\documentclass[aps,prb,reprint,groupedaddress,showpacs]{revtex4-1}

\usepackage{graphicx}
\usepackage{tikz}
\usepackage[caption=false]{subfig}
\usepackage{amsmath}
\usepackage{amssymb}
\usepackage{braket}
\usepackage{xfrac}
\usepackage{url}
\usepackage{filecontents}

\makeatletter
\newcommand{\colorcaption}[2][]{%
  \begingroup%
  \renewcommand{\@caption@fignum@sep}{ (Color online). }%
  \caption[#1]{#2}%
  \endgroup%
}
\makeatother

\renewcommand{\vec}[1]{\boldsymbol{#1}}
\newcommand{\efield}{\vec{\mathcal{E}}}
\newcommand\myatop[2]{\genfrac{}{}{0pt}{}{#1}{#2}}

\begin{document}

\title{First-principles approach to electric polarization and dielectric constant\\calculations using generalized Wannier functions}

\author{Pawel Lenarczyk}
\email{pawell@iis.ee.ethz.ch}
\author{Mathieu Luisier}
\affiliation{Integrated Systems Laboratory, ETH Z\"{u}rich, 8092 Z\"{u}rich, Switzerland}

\date{\today}

\begin{abstract}
We describe a method to calculate the electronic properties of an insulator under an applied electric field.
It is based on the minimization of an electric enthalpy functional with respect to the orbitals, which behave
as Wannier functions under crystal translations, but are not necessarily orthogonal.
This paper extends the approach of Nunes and Vanderbilt (NV) {[Phys.~Rev.~Lett.~\textbf{73}, 712 (1994)]},
who demonstrated that a Wannier function representation can be used to study insulating crystals in the presence of
a finite electric field. According to a study by Fern\'{a}ndez \emph{et~al.} {[Phys.~Rev.~B.~\textbf{58}, R7480 (1998)]},
first-principles implementations of the NV approach suffer from the impact of the localization constraint on the
orthogonal wave functions, what affects the accuracy of the physical results. We show that because non-orthogonal
generalized Wannier functions can be more localized than their orthogonal counterparts, the error due to localization
constraints is reduced, thus improving the accuracy of the calculated physical quantities.
\end{abstract}


\maketitle

\section{Introduction\label{sec_intro}}
The ability to perform first-principles calculations of solids under the influence of finite
electric fields is of fundamental as well as practical interest. Linear and nonlinear
susceptibilities of materials could be simultaneously extracted from such calculations
to determine their dielectric and ferroelectric behavior. These parameters can be
further used, for example, in the simulation of electronic devices\cite{ref_myFeFET}.

At this point the study of materials in a finite electric field remains a challenging
theoretical problem\cite{ref_fieldTheory}. The main difficulty comes from the scalar potential
``$\efield \cdot \vec{r}$'' that accounts for the electric field $\efield$. It induces a linear term
in the spatial coordinates $\vec{r}$ and thus violates the periodicity condition underlying
Bloch's theorem. This term acts as a singular perturbation to the electronic eigenstates.
As a consequence, standard computational methods relying on the solution of the eigenfunctions
of the effective one-electron Hamiltonian are not suitable for this kind of applications.

This restriction can be alleviated by making use of linear-response theory\cite{ref_LRT},
which provides a framework for computing derivatives of various quantities with respect to the
applied field. Practically, however, with such techniques only the response to infinitesimal
electric fields can be accurately studied. Moreover, their extension to nonlinear order is
not straightforward and must be carefully handled to avoid divergences in the
static limit\cite{ref_LRTdiv}.

In Ref.~\onlinecite{ref_NV}, Nunes and Vanderbilt (NV) proposed an approach to circumvent the difficulties
associated with finite electric fields. They showed that a real-space Wannier function
representation could be used to describe an insulating periodic system in the presence of
a finite electric field. In this scheme, an electronic enthalpy functional
$W$ is minimized with respect to localized orthogonal orbitals. The functional $W$ is made of
the usual band-structure energy $E_{bs}$ and a field coupling term $-\vec{P}_{el} \cdot \efield$.
Here $E_{bs}$ and the macroscopic electronic polarization $\vec{P}_{el}$ are expressed in terms
of Wannier functions (WFs), the latter via the modern theory of polarization\cite{ref_MTP}.
The NV approach was implemented within density functional theory by Fern\'{a}ndez \emph{et~al.}\cite{ref_polWf},
but was hindered by convergence problems with respect to the size of the localization regions of the truncated WFs.

In this paper, we propose an original formalism to perform first-principles calculations
of insulators under finite electric fields using non-orthogonal generalized Wannier functions (NGWFs).
This extension is motivated by the fact that non-orthogonal wave functions can be considerably more
localized than orthogonal ones\cite{ref_AndersonFunc}. Non-orthogonal orbitals have been
used in previous computational studies\cite{ref_orbSilicon,ref_DM_ON,ref_NGWF} at zero electric field.
In this case, they are not truly localized, but rather represent Bloch functions of the cell or supercell
calculated at a momentum $\vec{k}=0$. Hence they cannot be applied to finite field situations.
Instead, we employ here the orbitals that are truly localized in the manner of Wannier functions and explore
the effect of non-orthogonality also in the finite field case. This is done through the implementation of
a self-consistent scheme based on the  minimization of the electronic enthalpy functional expressed
in terms of NGWFs. The resulting solver is then utilized to study the ability of NGWFs to predict
the electronic and dielectric properties of materials from first-principles and to discuss the convergence of
these physical quantities with respect to the size of the orbital localization region.

The reminder of this paper is organized as follows. In the next section the formalism
is presented and its theoretical foundations are introduced. Details about the implementation
are given in Sec. \ref{sec_compdetails}, including a discussion of the minimization procedure and
a description of the calculations in real-space. In Sec. \ref{sec_results}, we show tests that
have been performed to probe the practical usefulness of the method and compare the calculation
results using localized wave functions with and without the orthogonality constraint.
Finally, in Sec. \ref{sec_conclusions} we conclude and mention possible future developments.

\section{Formalism\label{sec_formalism}}
In this work, the pseudopotential approximation\cite{ref_TMpseudo} to the Kohn-Sham
density functional theory\cite{ref_KS} (DFT) is employed. In this context, the problem of
interacting electrons and ions is mapped onto a problem of an effective system of $N$
non-interacting valence electrons following the potential of ions screened by the core electrons.
The effective Hamiltonian in real-space is equal to
\begin{equation}
   H[\rho](\vec{r}) = -\frac{1}{2}\nabla^2 + V_{ion}(\vec{r}) + V_{H}[\rho](\vec{r}) + V_{xc}[\rho](\vec{r}) ~.
   \label{eq_Hr}
\end{equation}
Here, the notation $H[\rho](\vec{r})$ indicates that the Hamiltonian $H$ is a functional of
the electron density $\rho(\vec{r})$ and a function of the spatial coordinate $\vec{r}$.
Expressed in the same form, $V_{ion}$ is the ion core pseudopotential, $V_{H}$ the Hartree
potential, and $V_{xc}$ the exchange-correlation potential. In this work, the local density
approximation (LDA) is adopted and $V_{xc}[\rho](\vec{r})$ is derived from the value of
the charge density, i.e. $V_{xc}[\rho](\vec{r})=V_{xc}(\rho(\vec{r}))$.
Atomic units $\bar{e}=\hbar=m_{e}=1$ are used throughout the paper.

Within DFT, the total energy of a system of interacting electrons and ions
is a unique functional of the electron density $\rho(\vec{r})$ and can be written as
\begin{equation}
 E[\rho] = E_{bs}[\rho] - E_{dc}[\rho] + E_{II} ~,
 \label{eq_Etot}
\end{equation}
where $E_{bs}$ is the band-structure energy that is defined as a trace of the
Hamiltonian in the occupied space,
$E_{dc}=\int \rho(\vec{r}) \left( \frac{1}{2}V_{H}(\vec{r}) + V_{xc}(\rho(\vec{r})) - \epsilon_{xc}(\rho(\vec{r})) \right) \mathrm{d}^3\vec{r}$
accounts for double counting in the Coulomb electronic repulsion and for
the exchange-correlation corrections, while $E_{II}$ is the Coulomb
interaction energy between the ions. By applying the variational principle of
Hohenberg and Kohn\cite{ref_HK}, the ground state energy of the system can be obtained.
This requires minimizing the total energy functional $E$ in Eq.~(\ref{eq_Etot})
with respect to either single particle wave functions\cite{ref_iterMin}
or density matrices\cite{ref_DM_ON}.

In the density matrix description, the expectation value of any operator
$\hat{O}$ is given by $\mathrm{tr} [ \hat{O} \hat{\rho} ]$,
where $\hat{\rho}$ is the density matrix operator defined as the
projection onto the occupied space. The diagonal element of the density matrix
in a spatial representation corresponds to the charge density
$\rho(\vec{r})= 2 \bra{\vec{r}} \hat{\rho} \ket{\vec{r}}$,
where the factor~of~$2$~accounts for the spin degeneracy.
In this formalism the band-structure energy is given by
\begin{equation}
 E_{bs}[\rho] = 2 \mathrm{tr} \left[ \hat{\rho} \hat{H} \right] ~.
 \label{eq_Ebs}
\end{equation}
In Eq.~(\ref{eq_Ebs}), $\hat{H}$ is the Hamiltonian operator, in the position
representation given by Eq.~(\ref{eq_Hr}).
Minimizing Eq.~(\ref{eq_Ebs}) for a fixed, density-independent Hamiltonian and
finding a self-consistent solution for the charge density is equivalent to solving
the non-linear Kohn-Sham equations\cite{ref_iterMin}.

Our goal is to perform the calculations at finite electric fields.
The Hamiltonian then becomes
\begin{equation}
 H[\rho](\efield;\vec{r}) = H[\rho](\vec{r}) + \efield \cdot \vec{r} ~.
 \label{eq_Hfield}
\end{equation}
It has now a parametric dependence on the electric field $\efield$.
Replacing $H[\rho](\vec{r})$ by $H[\rho](\efield;\vec{r})$ in Eq.~(\ref{eq_Ebs})
leads to the electronic enthalpy functional
\begin{equation}
 W[\rho](\efield) = E_{bs}[\rho] - \Omega \efield \cdot \vec{P}_{el}[\rho] ~,
 \label{eq_Wdef}
\end{equation}
whose minimization with respect to a set of field-dependent density matrices
$\hat{\rho}(\efield)$ results in the electronic ground state of an insulator
in the presence of an electric field. The variable $\vec{P}_{el}[\rho]$
refers to the electronic macroscopic polarization. In the density matrix
formalism, it is defined as
\begin{equation}
 \vec{P}_{el}[\rho] = -\frac{2}{\Omega} \mathrm{tr} \left[ \hat{\rho} \hat{\vec{r}} \right] ~,
 \label{eq_Pel_def}
\end{equation}
where $\hat{\vec{r}}$ is the position operator and $\Omega$ is the volume of
the chosen unit cell.

For an insulating crystal, the density operator $\hat{\rho}$ may be expanded in terms of
localized functions $\{\nu_a^i\}$, represented below in Dirac's bra-ket notation, as\cite{ref_WFrev}
\begin{equation}
 \hat{\rho} = \sum_{\myatop{ij}{ab}} \ket{\nu_a^i} K_{ab}^{ij} \bra{\nu_b^j} ~.
 \label{eq_densop_param}
\end{equation}
The upper-case sum over $i,j$ runs over cell replicas whereas the lower-case sum over $a,b$ goes over occupied bands.
The periodicity of the crystal is taken into account by imposing that any wave function is obtained by translating
that of a reference unit cell, denoted by index $0$, with the help of Bravais lattice vectors
\begin{equation}
 \ket{\nu_a^i} = \hat{T}_{\vec{R}_i} \ket{\nu_a^0} ~,
 \label{eq_wf_period}
\end{equation}
where $\hat{T}_{\vec{R}_i}$ is the translation operator corresponding to the lattice vector $\vec{R}_i$.
For the \emph{density kernel} matrix $\mathbf{K}$ the following relation holds\cite{ref_WFrev}
\begin{equation}
 K_{ab}^{ij}=K_{ab}(\vec{R}_j - \vec{R}_i) ~.
 \label{eq_Kmat_period}
\end{equation}

In the above parametrization of the density operator in Eq.~(\ref{eq_densop_param}),
the $\mathbf{K}$ matrix plays the role of an inverse overlap matrix between the
generally non-orthogonal $\{\nu_a^i\}$ functions. These functions are called
non-orthogonal generalized Wannier functions (NGWFs).
Note that for $K_{ab}^{ij}=\delta_{ij}\delta_{ab}$, the wave functions are orthogonal and
correspond to the standard Wannier functions (WFs) of a periodic system.
For a discussion of the ground state $\mathbf{K}$~matrix properties see Appendix \ref{sec_app_chempot}.

It should be emphasized that similar parametrizations of the density matrix in terms of
non-orthogonal orbitals as in Eq.~(\ref{eq_densop_param}) were proposed and used by a number
of other authors\cite{ref_NGWF,ref_DM_ON,ref_DMparam,ref_ONmetals}. However, in all these
studies, the sum over periodic replicas was dropped and the calculations were performed
at the $\Gamma$-point only, i.e. $\vec{k}=0$. The wave functions employed in these
investigations are in fact extended Bloch functions of the solid. They cannot be directly
used to study the response of a periodic system to an electric field because of
ill-posedness of the position operator $\hat{\vec{r}}$ in the Bloch representation,
as explained in Appendix~\ref{sec_app_posop}.

As next step, Eq.~(\ref{eq_densop_param}) is taken as an ansatz for a \emph{trial} density operator
and the \emph{physical} density operator
\begin{equation}
 \hat{\rho}' = 2 \hat{\rho} - \hat{\rho}^2
 \label{eq_densop_transf}
\end{equation}
is introduced. The above \emph{purifying} transformation\cite{ref_McWeeny} ensures that $\hat{\rho}'$ does not have
eigenvalues larger than $1$. This condition on the eigenvalues is termed \emph{weak idempotency}.
It is critical to give the underlying energy functional the desired minimal properties\cite{[{See }][{, and the references therein for details.}]ref_ONrev}.

In the chosen formalism the expectation value of any operator $\hat{O}$ is re-expressed as $\mathrm{tr}[\hat{\rho}' \hat{O}]$.
Thus, the band-structure energy and the electronic polarization per unit cell become
\begin{equation}
 E_{bs} = 2 \sum_{\myatop{i}{ab}} Q_{ab}^{0i} \bra{\nu_a^0} \hat{H} \ket{\nu_b^i}
 \label{eq_Ebs_expr}
\end{equation}
and
\begin{equation}
 \vec{P}_{el} = - \frac{2}{\Omega} \sum_{\myatop{i}{ab}} Q_{ab}^{0i} \bra{\nu_a^0} \hat{\vec{r}} \ket{\nu_b^i} ~,
 \label{eq_Pel_expr}
\end{equation}
where
\begin{equation}
 Q_{ab}^{ij} = 2 K_{ab}^{ij} - \left( \mathbf{K} \times \mathbf{S} \times \mathbf{K} \right)_{ab}^{ij}
 \label{eq_Qmat_def}
\end{equation}
and $\mathbf{S}$ is the overlap matrix between the orbitals
\begin{equation}
 S_{ab}^{ij} = \braket{\nu_a^i | \nu_b^j} ~.
 \label{eq_Smat}
\end{equation}
In Eq.~(\ref{eq_Qmat_def}) the short-hand notation
\begin{equation}
 \left( M \times N \right)_{ab}^{ij} = \sum_{\myatop{k}{c}} M_{ac}^{ik} N_{cb}^{kj}
 \nonumber
\end{equation}
is introduced to represent matrix-matrix multiplications between overlap-type matrices
$M_{ac}^{ij}=M_{ac}(\vec{R}_j - \vec{R}_i)$.

The charge density is then given by
\begin{equation}
 \rho(\vec{r}) = 2 \sum_{\myatop{ij}{ab}}  \braket{\vec{r}|\nu_a^i} Q_{ab}^{ij} \braket{\nu_b^j|\vec{r}} ~.
 \label{eq_dens_expr}
\end{equation}

Substituting Eqs.~(\ref{eq_Ebs_expr})~and~(\ref{eq_Pel_expr}) into Eq.~(\ref{eq_Wdef})
leads to an expression for the electronic enthalpy per unit cell
\begin{equation}
 \begin{split}
   W = 2\sum_{\myatop{i}{ab}} & \big( 2 K_{ab}^{0i} - \sum_{\myatop{jk}{cd}} K_{ac}^{0j} \braket{\nu_c^j | \nu_d^k} K_{db}^{ki} \big) \times \\
                             & \times \bra{\nu_a^0} \hat{H}(\efield) \ket{\nu_b^i}  ~,
 \end{split}
 \label{eq_W_expr}
\end{equation}
where $\hat{H}(\efield) = \hat{H} + \efield \cdot \hat{\vec{r}}$.
We note that for $K_{ab}^{ij}=\delta_{ij}\delta_{ab}$ the functional in Eq.~(\ref{eq_W_expr})
corresponds to the one originally proposed by NV, which is minimized by orthogonal orbitals.
The introduction of the $\mathbf{K}$ matrix in our approach gives more variational freedom
for the minimization and results in non-orthogonal orbitals that can be made more localized,
as it will be shown in the following.

The electronic degrees of freedom are the coefficients of the wave functions $\{\nu_a^0\}$
and the density kernel matrix elements $\{K_{ab}^{0i}\}$ in the base cell corresponding to $\vec{R}_0=0$.
The remaining variables are obtained by employing the periodicity relations in
Eqs.~(\ref{eq_wf_period})~and~(\ref{eq_Kmat_period}) for the wave functions and density kernel matrix elements,
respectively. Using these conditions, the extermized functional in Eq.~(\ref{eq_W_expr}) can be
written explicitly in terms of minimization variables as
\begin{equation}
 \begin{split}
   W = 2\sum_{\myatop{i}{ab}} & \big( 2 K_{ab}^{0i} - \sum_{\myatop{jk}{cd}} K_{ac}^{0j} \braket{\nu_c^0 | \nu_d^{k-j}} K_{db}^{0i-k} \big) \times \\
                             & \times \bra{\nu_a^0} \hat{H}(\efield) \big( \hat{T}_{\vec{R}_i} \ket{\nu_b^0} \big) ~.
 \end{split}
 \label{eq_W_comp}
\end{equation}

The search for the minimum of the functional $W$ requires the knowledge of its partial derivatives with
respect to the variational degrees of freedom. In the above Eq.~(\ref{eq_W_comp}), the partial
derivatives of $W$ with respect to the minimization variables $\{\nu_a^0\}$ and $\{K_{ab}^{0i}\}$
can be carried out. Differentiating Eq.~(\ref{eq_W_comp}) with respect to $\nu_a^0$ gives:
\begin{equation}
 \begin{split}
  \frac{\partial W}{\partial \nu_a^0} = 4 \sum_{\myatop{i}{b}} & \bigg[ \hat{H}(\efield) \ket{\nu_b^i} Q_{ab}^{0i} + \\
  & - \ket{\nu_b^i}  \big( \mathbf{K} \times  \mathbf{H}(\efield) \times \mathbf{K} \big)_{ab}^{0i} \bigg] ~.
 \end{split}
 \label{eq_dWdwf}
\end{equation}
The partial derivative of $W$ with respect to $K_{ab}^{0i}$ has the following form:
\begin{equation}
 \begin{split}
 \frac{\partial W}{\partial K_{ab}^{0i}} = 2 \bigg[ & 2H_{ab}^{0i} -
 \big( \mathbf{H}(\efield) \times \mathbf{K} \times \mathbf{S} \big)_{ab}^{0i} + \\
 & - \big( \mathbf{S} \times \mathbf{K} \times \mathbf{H}(\efield) \big)_{ab}^{0i} \bigg]
 \end{split}
 \label{eq_dWdK}
\end{equation}
Above, $\mathbf{H}(\efield)$ stands for the matrix representation of the Hamiltonian
$H_{ab}^{ij}=\bra{\nu_a^i} \hat{H} \ket{\nu_b^j}$
in the basis of the employed localized orbitals, WFs or NGWFs.

\section{Computational details\label{sec_compdetails}}
We now give a brief description of our implementation of the formalism presented above.
It was integrated into \texttt{\MakeUppercase{parsec}} \cite{ref_PARSEC_web} open-source DFT code.
The modified package was used to obtain the results reported in the following section.

The wave functions $\{v_a^0\}$ are represented on a uniform real-space grid
with spacing $h$ in each direction. Since these functions are required to be spatially
localized, they have non-zero values only on the grid points inside the localization
regions (LRs). In the present work, we let each LR be a cube of edge size $a_{LR}$.
The centers of the LRs may be chosen arbitrarily. For the cutoff of the density kernel matrix,
the $K_{ab}^{0i}$ elements are non-zero only if the LRs of $v_a^0$ and $v_b^i$ overlap.
Note that if this localization condition is imposed, the sums over periodic replicas
($i$,$j$,$k$,$l$) appearing in the preceding section become finite. They are determined
by the set of LRs that overlap with all LRs centered in the supercell containing the
origin and indicated by the index 0.

By using a homogeneous grid, the real-space integration is replaced by a summation over the
discretization points, so that, e.g., the overlap matrix elements can be calculated as
\begin{equation}
 S_{ab}^{ij} \simeq h^3 \sum_{\vec{r}_g} v_a^i(\vec{r}_g) v_b^j(\vec{r}_g) ~.
 \nonumber
\end{equation}
The sum goes over the set of grid points $\vec{r}_g$ that are shared by the localization regions
of both $v_a^i$ and $v_b^j$ orbitals. Note that $h^3$ is the volume of each grid point.

The Hamiltonian operator is evaluated directly on the real-space grid, as implemented in the
\texttt{\MakeUppercase{parsec}} code\cite{ref_PARSEC_FD}. A finite-difference expansion of order $M$
of the Laplacian $\nabla^2$ is used to evaluate the kinetic energy operator. The ionic potential
$V_{ion}$ is determined by a pseudopotential cast in the Kleinman-Bylander form\cite{ref_KBpseudo}.
The Hartree and exchange-correlation potentials $V_H(\vec{r})$ and $V_{xc}(\vec{r})$ are represented
by numerical values on the grid. The real-space Hamiltonian $H(\vec{r})$ in \texttt{\MakeUppercase{parsec}}
is defined only in the base supercell. To act on the localized orbitals $\{v_a^i\}$ the Hamiltonian 
is circularly shifted to the LRs of the orbitals and evaluated there.

In the tests presented below, we use norm-conserving pseudopotentials generated with the method of
\mbox{Troullier} and \mbox{Martins}\cite{ref_TMpseudo} and obtained from Ref.~\onlinecite{ref_PARSEC_web}.
The exchange and correlation effects are treated within the local-density functional of
Ceperley and Alder\cite{ref_LDA_CA}, as parameterized by Perdew and Zunger\cite{ref_LDA_PZ}.

The evaluation of the Hamiltonian and two-center integrals allows one to calculate the enthalpy
functional $W$ in Eq.~(\ref{eq_W_comp}) and its derivatives $\frac{\partial W}{\partial \nu_a^0}$
and $\frac{\partial W}{\partial K_{ab}^{0i}}$ in Eqs.~(\ref{eq_dWdwf})~and~(\ref{eq_dWdK}),
respectively. It enables a search for the ground state that minimizes the energy.
This optimization can be carried out in two nested stages
\begin{equation}
 W_{min} = \min_{ \small { \{v_a^0\} } } W'(\{v_a^0\}) ~,
 \label{eq_min_wf}
\end{equation}
with
\begin{equation}
 W'(\{v_a^0\}) = \min_{ \small { \{K_{ab}^{0i}\} } } W(\{K_{ab}^{0i}\},\{v_a^0\}) ~.
 \label{eq_min_kern}
\end{equation}
The minimization with respect to the density kernel in Eq.~(\ref{eq_min_kern}) ensures that
$W'$ of Eq.~(\ref{eq_min_wf}) is a function of NGWFs only.
The above minimizations has been implemented with a conjugate gradient scheme based on the
analytical gradients from Eq.~(\ref{eq_dWdwf}) to optimize the orbitals in Eq.~(\ref{eq_min_wf})
and from Eq.~(\ref{eq_dWdK}) for the optimization of the density kernel in Eq.~(\ref{eq_min_kern}).
The gradients are made mutually conjugate using Polak-Ribi\`{e}re formula\cite{ref_CG_PR}.
The nested minimization approach is inspired by the method developed in Ref.~\onlinecite{ref_ONETEP_Kmin}
in the context of zero field calculations with periodic non-orthogonal wave functions.
When optimizing WFs we set the density kernel to an identity matrix $K_{ab}^{0i}=\delta_{0i}\delta_{ab}$
and skip the minimization in Eq.~(\ref{eq_min_kern}). This results in the orthogonal
wave functions as shown in Appendix~\ref{sec_app_chempot}.

During the electronic enthalpy minimization, as described above, the Hamiltonian is kept fixed.
This has the practical advantage that the enthalpy functional in Eq.~(\ref{eq_W_comp}) is a
quartic function of the $\{v^a_0\}$ coefficients and a quadratic function of the $\{K_{ab}^{0i}\}$ elements.
The conjugate-gradient line searches can therefore be solved exactly\cite{ref_NumRecipes} by computing the
coefficients of the fourth and second-order polynomials for the orbital and density kernel optimizations,
respectively. In order to assure the existence of a minimum, the Hamiltonian eigenspectrum is shifted by a
transformation $H \rightarrow H - \mu I$, where $\mu$ is a free parameter that makes all the eigenvalues
negative. A discussion of this transformation and its impact on the minimized enthalpy functional
is derefered to Appendix~\ref{sec_app_chempot}.

The results of the minimization algorithm are the grid coefficients $\{v_a^0(\vec{r}_g)\}$ of the wave functions
in the base cell and the corresponding matrix elements $\{K_{ab}^{0i}\}$. The functions $\{v_a^i\}$ and the elements
$\{K_{ab}^{ij}\}$ are evaluated \mbox{on-the-fly} when calculating the sums over the replicas ($i$,$j$,$k$,$l$). The periodicity
relations of the wave functions in Eq.~(\ref{eq_wf_period}) and of the matrix elements in Eq.~(\ref{eq_Kmat_period})
are exploited to do that. The charge density $\rho(\vec{r})$ is periodic in the supercell and is calculated according
to Eq.~(\ref{eq_dens_expr}). At the end of the minimization procedure, if it is found that the charge density
as well as the Hartree and exchange-correlation potentials are not consistent, the whole operation is repeated with the
updated potentials, as in standard self-consistent field (SCF) cycle.

\section{Results\label{sec_results}}

In this section the application of the above described methodology is presented.
In particular, we emphasize that the main purpose of the conducted study is to exhibit
and understand the impact of the localization constraint on the accuracy of
the ground state and finite-field calculations, using orthogonal and non-orthogonal
Wannier functions: WFs and NGWFs respectively.

Since DFT is variational\cite{ref_elecSt}, any restriction placed on the class of density matrices
that can be searched over has the effect of raising the minimum energy $E_{bs}$ Eq.~(\ref{eq_Ebs})
above its true ground state value $E_{0}$. This suggests that the error introduced by using
LRs of finite size $a_{LR}$ in the minimization Eq.~(\ref{eq_min_wf}) can be assessed by calculating
\begin{equation}
 \Delta E_{bs}(a_{LR}) = E_{bs}(a_{LR}) - E_{0} ~,
 \label{Eq_dEbs}
\end{equation}
where $E_{bs}(a_{LR})$ is the minimum band-structure energy at $a_{LR}$ and $E_{0}$ is the reference energy,
with no localization constraints. The value of $E_{0}$ can be estimated by the conventional diagonalization
of the Hamiltonian using Bloch functions and converged $\vec{k}$-point sampling.

In general it can be presumed that the density matrix $\rho(\vec{r},\vec{r}')$ in the true
ground state tends to zero as the separation of its arguments $|\vec{r}-\vec{r}'|$ increases \cite{ref_projAnalytic}.
In~an~early pioneering work, Kohn\cite{ref_expDecay_Kohn} proved that the density matrix and
Wannier functions for a one-dimensional (1D) model crystal decay exponentially
in systems with a band~gap.
In a more recent work, He and Vanderbilt \cite{ref_expDecay_HV} demonstrated that in 1D insulators
the spatial decay of the density matrix and Wannier functions has the form of a power
law times exponential, what results in a faster decay than that predicted
by Kohn. These authors have also shown that for 1D model problems
non-orthogonal Wannier-like functions exhibit superior localization as compared to
orthogonal ones.
In the all three spatial dimensions, the exponential decay of the Wannier functions
has been proven for a single-band case \cite{ref_expLocWF}, and that of the density
matrix has been proven in general \cite{ref_projAnalytic}. For the density matrix also
some simple predictions of the inverse decay length are available,
in the tight-binding \cite{ref_Kohn_DFWFT}, and weak-binding \cite{ref_locDM_weakBind} limits.

The above considerations strongly suggests that the the error in the energy should go quickly to zero
with the increase of the real-space cutoff $R_C$ imposed~on~$\rho(\vec{r},\vec{r}')$:
$\rho(\vec{r},\vec{r}')=0,~ |\vec{r}-\vec{r}'|>R_C$.
The localization properties of the density matrix create a fundamental basis for the
development of various expansion algorithms
\cite{ref_DMalg1,ref_DMalg2,ref_DMalg3,ref_DMalg4,ref_DM_LNV,ref_DM_NV}
which enable calculations of density matrix with computational complexity that scale linearly with system size.
In our formulation the cutoff $R_C$ is controlled by the size of LRs, $\frac{a_{LR}}{2}$, of the localized
functions $\{\nu_a^i\}$ and the dimension of the density kernel matrix $\mathbf{K}$, as can be seen by substituting
$\rho(\vec{r},\vec{r}')=\bra{\vec{r}} \hat{\rho} \ket{\vec{r}'}$ in the expansion Eq.~(\ref{eq_densop_param}).
It will be verified by the test calculations below that the error $\Delta E_{bs}(a_{LR})$ decreases with
increasing $a_{LR}$.

A strict localization is not compatible with orthogonality\cite{ref_AndersonFunc}. This introduces an error
in the total particle number, what can be examined by looking at the quantity
\begin{equation}
 \Delta{}N(a_{LR}) = N - \int{\rho(a_{LR};\vec{r})~\mathrm{d}\vec{r}} ~,
 \label{Eq_dNtot}
\end{equation}
where $N$ is the number of valence electrons in the system, and $\rho(a_{LR};\vec{r})$ denotes
the charge density optimized at $a_{LR}$. Since there are only $\frac{N}{2}$ eigenvalues of $\hat{\rho}$
that are different from $0$, and because these eigenvalues are constrained to be smaller or equal to $1$,
$N$ is a rigorous upper bound to $2 \mathrm{tr} \hat{\rho} = \int{\rho(\vec{r})~\mathrm{d}\vec{r}}$
and $\Delta{}N$ is necessarily non-negative as it will be exemplified below.

The electronic response of an insulating solid to applied electric field $\efield$ can be quantified
by considering high-frequency dielectric constants. The dielectric tensor $\mathbf{\epsilon}$
is related to the macroscopic polarization $\vec{P}$ by
\begin{equation}
 \epsilon_{\alpha{}\beta} = \delta_{\alpha{}\beta} +
 4\pi \frac{\mathrm{d}P_{\alpha}}{\mathrm{d}\mathcal{E}_{\beta}} ~,
 \label{Eq_eps}
\end{equation}
where $\alpha$ and $\beta$ indicate Cartesian coordinates. It can be obtained by finite differences
of the induced polarization vector components $P_{\alpha}$ for different values of $\mathcal{E}_{\beta}$
coefficients. That is, $\mathcal{E}_{\beta}$ is increased by small increments $\Delta{}\mathcal{E}$
and the computed values of $P_{\alpha}$ are used to evaluate 1-st order approximation:
$\frac{\mathrm{d}P_{\alpha}}{\mathrm{d}\mathcal{E}_{\beta}} \approx
 \frac{P_{\alpha}(\mathcal{E}_{\beta}+\Delta{}\mathcal{E})-P_{\alpha}(\mathcal{E}_{\beta})}{\Delta{}\mathcal{E}}$
, substituted in Eq.~(\ref{Eq_eps}). The calculation is repeated for a few values of $\Delta{}\mathcal{E}$
to minimize the impact of numerical errors.
If the ions are kept fixed, as for the results reported below, this gives high-frequency dielectric
constants $\epsilon_{\alpha{}\beta}^{\infty}$. It will be shown that the calculation of
$\mathbf{\epsilon}^{\infty}$ converges exponentially as the size of the localization $a_{LR}$
is increased. Fitting the calculated values of $\epsilon^{\infty}$ at different $a_{LR}$,
using the function
\begin{equation}
 \epsilon^{\infty}(a_{LR})=A~\mathrm{exp}(-(a_{LR}/a)/\alpha)+B ~,
 \label{Eq_epsfit}
\end{equation}
allows to extract dielectric constants in the limit of no localization constraint
$\epsilon^{\infty}(a_{LR}\rightarrow{}\infty)=B$.

Two different systems were selected to test the implementation of our method.
The first structure --- bulk Silicon is chosen due to its practical relevance and
because it often serves as a benchmark for DFT codes\cite{ref_DM_ON,ref_orbSilicon,ref_polWf}.
The second one is cubic
$\mathrm{BaTiO_{3}}$ which has very interesting bonding properties examined so far only using
Maximally localized Wannier functions\cite{ref_WFrev} (MLWFs), constructed from Bloch orbitals
at zero electric field\cite{ref_btoWf}.

\subsection{Bulk Silicon\label{sec_Si}}

The simulation of bulk Silicon is performed using periodically continued cubic cell
containing $8$ atoms arranged in a diamond structure. The lattice parameter used is
$a=10.2~\mathrm{a_{0}}$, $\mathrm{a_{0}}$ denoting the atomic length unit.
A total of $N=32$ valence electrons are accounted for by $16$ doubly occupied orbitals.
The LRs of the considered orbitals are centered on the bonds connecting one $\mathrm{Si}$
atom with its four nearest neighbors. Their start values are assumed to be centrosymmetric
Gaussian functions with their origin at the localization centers and  their variance corresponding to
half of the \mbox{$\mathrm{Si}$--$\mathrm{Si}$} bond length. The identity matrix is used
as a initial guess for $\mathbf{K}$. The converged grid spacing is $h=0.3~\mathrm{a_{0}}$
and we employ $M=6$ order finite difference expansion for the Laplacian operator. The chemical potential
is set to $\mu=3~\mathrm{Ry}$. With these parameters, the minimizations are carried out until
the change in the energy is less than $10^{-6}\mathrm{Ry/atom}$.
For comparison purposes, we also perform calculations using extended Bloch functions with the
same real-space setup as for the localized wave functions and $3\times3\times3$ Monkhorst-Pack grid\cite{ref_MPgrid}
for the computations in reciprocal space.

\begin{figure}[h]
\centering
\includegraphics[width=0.99\linewidth]{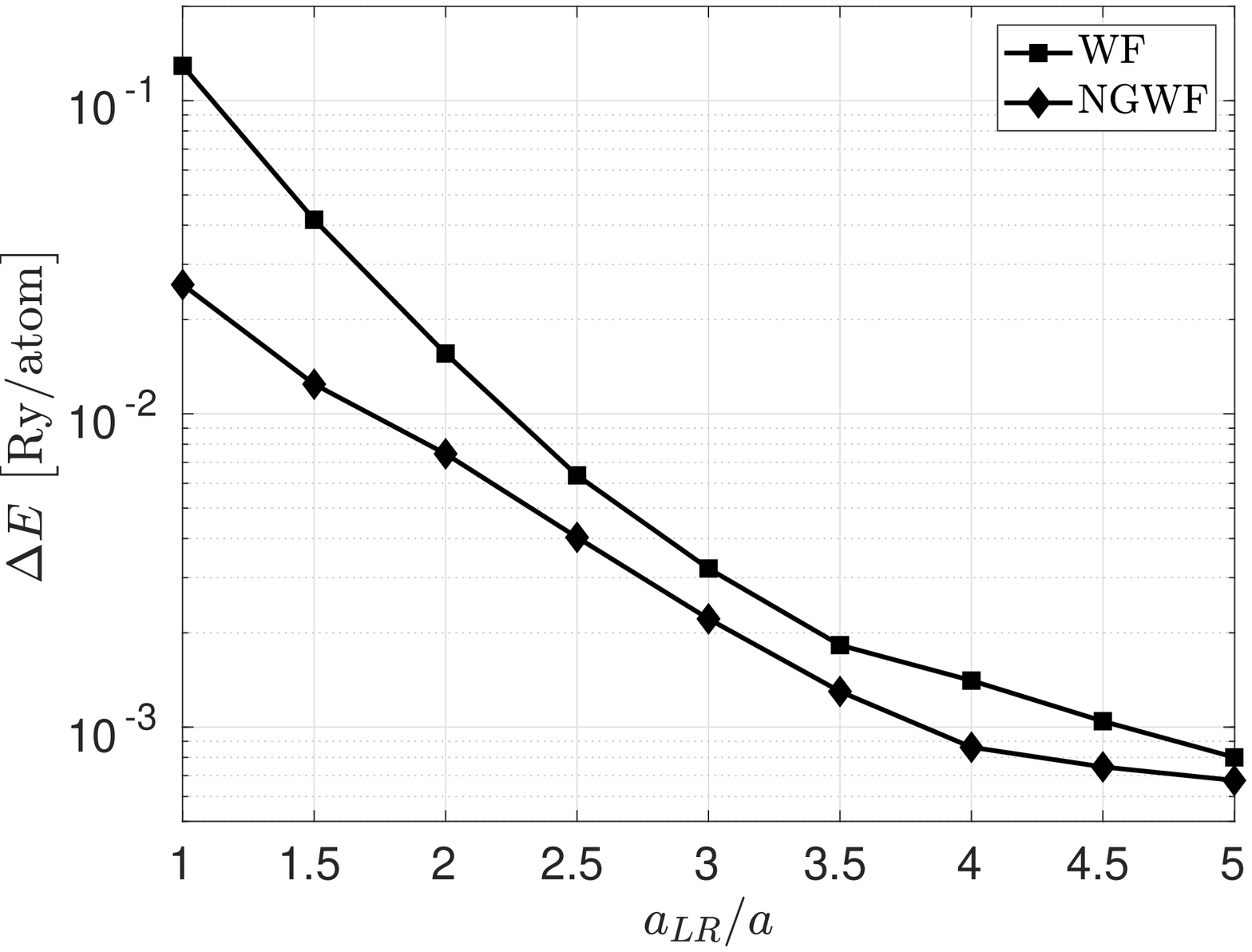}
\caption{$\mathrm{Si}$: Error in band-structure energy $\Delta{}E_{bs}$ versus the size $a_{LR}$ of the
localization regions. Squares: Wannier functions (WFs). Diamonds: non-orthogonal generalized Wannier functions (NGWFs).
$a_{LR}$ is normalized with the lattice constant $a=10.2\mathrm{a_{0}}$.
\label{fig_Si_dEbs}}
\end{figure}

Fig.~\ref{fig_Si_dEbs} shows the error in the band-structure energy from Eq.~(\ref{Eq_dEbs})
evaluated for different LR sizes $a_{LR}$ and compared to the reference value resulting from
a sum of $\frac{N}{2}$ doubly-occupied Bloch eigenstates over the considered $\vec{k}$-points.

\begin{figure}[t]
\centering
\includegraphics[width=0.99\linewidth]{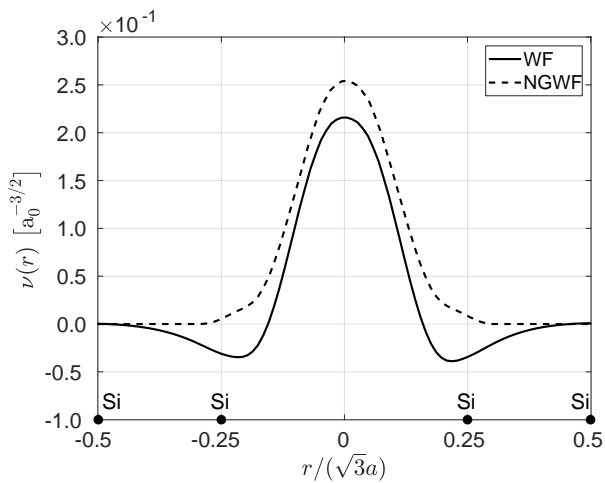}
\caption{$\mathrm{Si}$: Line plot of $\sigma$-oriented wave functions along \mbox{$\mathrm{Si}$--$\mathrm{Si}$} bond in the ground state.
$|r|$ is the distance from the localization center. Solid line: WF. Dashed line: NGWF.
The wave functions are constrained to be zero outside the localization region of size $a_{LR}=2a$.
\label{fig_Si_wfcut}}
\end{figure}

The results displayed in Fig.~\ref{fig_Si_dEbs} demonstrate that the effect of non-orthogonality is
especially pronounced for strict localization constraints: $a_{LR}<2a$. This improvement can be
understood by examining Fig.~\ref{fig_Si_wfcut} which plots the profiles of the wave functions along
\mbox{$\mathrm{Si}$--$\mathrm{Si}$} bond. As can be observed both WFs and NGWFs are localized, but in the
WF case the tail of the wave function does not decay as rapidly as with NGWF. This feature
of WF is dictated by the orthogonality requirement. The two roots of $\nu(r)$ in Fig.~\ref{fig_Si_wfcut}
correspond to the positions where the wave function must be zero in order to be orthogonal to the neighboring
orbitals. Consequently, the NGWFs can be made better localized than WFs, because they do not need to fulfill
the orthogonality constraints. This results into a reduction of the energy error $\Delta{}E_{bs}$
when working with non-orthogonal orbitals.

\begin{figure}[h]
\centering
\includegraphics[width=0.99\linewidth]{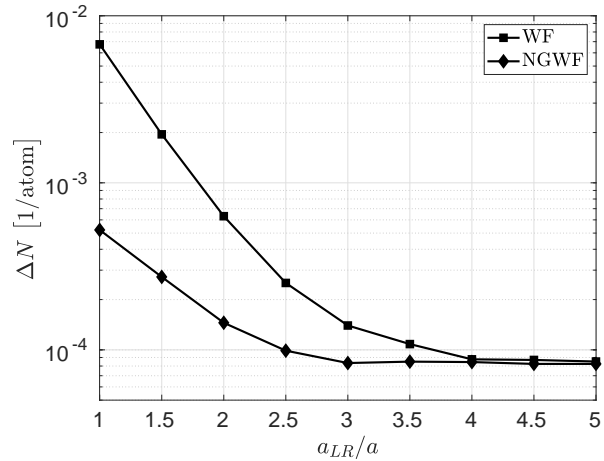}
\caption{$\mathrm{Si}$: Error in total particle number $\Delta{}N$ versus localization size $a_{LR}$.
Squares: WFs. Diamonds: NGWFs.
\label{fig_Si_dNtot}}
\end{figure}

Fig.~\ref{fig_Si_wfcut} confirms that strict localization is not compatible with orthogonality.
The corresponding error in total particle number from Eq.~(\ref{Eq_dNtot}) is plotted in
Fig.~\ref{fig_Si_dNtot} as a function of the localization size $a_{LR}$. The results show that
the accuracy losses due to the orthogonality constraint in the limit of strong localization can be
significantly reduced by using NGWFs instead of WFs. This enhancement can be attributed to the
inclusion of the density kernel matrix $\mathbf{K}$ to compensate for the non-orthogonality of the orbitals.

\begin{figure}[h]
\centering
\includegraphics[width=0.99\linewidth]{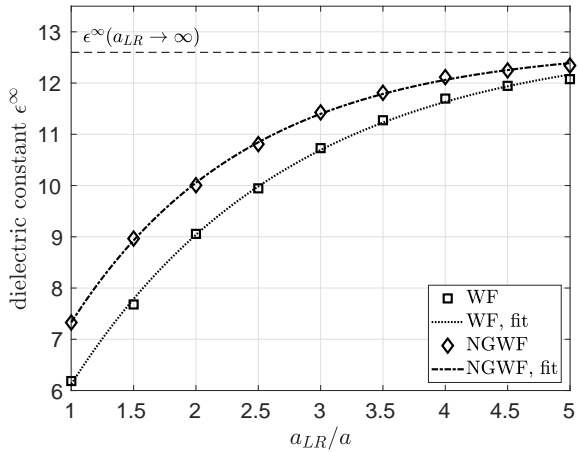}
\caption{$\mathrm{Si}$: Electronic dielectric constant $\epsilon^{\infty}$ computed with WFs (squares) and NGWFs (diamonds)
for different sizes of the localization regions $a_{LR}$. The dotted and dash-dotted lines display fittings of Eq.~(\ref{Eq_epsfit})
to WF and NGWF data, respectively. The vertical dashed line shows the average of the extrapolated fitted curves in the limit
$a_{LR}\rightarrow{}\infty$.
\label{fig_Si_eps}}
\end{figure}

As next step, we address the question of relaxing the orthogonality constraint on the localized
wave functions and its influence on the electronic response of bulk silicon to external electric fields.
In order to quantify the difference between WFs and NGWFs, the electronic dielectric constant $\epsilon^{\infty}$
has been calculated using Eq.~(\ref{Eq_eps}) at different $a_{LR}$. For the considered zinc blende structure it holds
$\epsilon=\epsilon_{xx}=\epsilon_{yy}=\epsilon_{zz}$.
The electric field used to induce the electronic response is applied along the [100] direction with maximum intensity
equal to $\bar{e}|\efield|=10^{-2}\mathrm{\sfrac{Ry}{a_{0}}}$. The weak intensity of the field guarantees the
linear response of the material and lies well below the onset of the Zener breakdown.
The results are shown in Fig.~\ref{fig_Si_eps}.

As can be seen in Fig.~\ref{fig_Si_eps} the convergence of $\epsilon^{\infty}$ with respect to
$a_{LR}$ is exponential. Fitting the data using the function in Eq.~(\ref{Eq_epsfit}) gives the limit of the no localization
constraint $a_{LR}\rightarrow{}\infty$. The value of $\epsilon^{\infty}(a_{LR}\rightarrow{}\infty)$ is $12.7$ for WFs
and $12.8$ for NGWFs. Fern\'{a}ndez \emph{et~al.} reported in Ref.~\onlinecite{ref_polWf} a value of $13.4$ for analogous calculations
using orthogonal Wannier functions. The LRT results of $\epsilon^{\infty}$ vary between $12.6$ to $12.9$, depending on the details
of the calculation\cite{[{See for instance }]ref_epsSiLRT_1,*[][{, and references therein.}]ref_epsSiLRT_2}.
Therefore, it can be concluded that the difference in the extracted values of $\epsilon^{\infty}$
between our results and those of Fern\'{a}ndez \emph{et~al.} remains within an acceptable range. The convergence of $\epsilon^{\infty}$
with respect to $a_{LR}$ was also found to be exponential in Ref.~\onlinecite{ref_polWf} for bulk $\mathrm{Si}$ using orthogonal Wannier functions.
Note that the definition of localization size in Ref.~\onlinecite{ref_polWf} corresponds to $\frac{1}{2}a_{LR}$ in our work.
The fitting parameters of Eq.~(\ref{Eq_eps}) for WF and NGWF data in Fig.~\ref{fig_Si_eps} as well as the results from Ref.~\onlinecite{ref_polWf}
are given in Table~\ref{tab_Si_epsfit}. As it can be seen our calculation with orthogonal Wannier functions converges slower than that of
Fern\'{a}ndez \emph{et~al.}. As a consequence large LRs are required for reliable estimations of the electronic response.
The situation is improved by using non-orthogonal orbitals. If, for example, a convergence to within $5\%$ of
$\epsilon^{\infty}(a_{LR}\rightarrow{}\infty)$ is acceptable, then a localization region with $a_{LR}=3.7a$ is needed in the case of NGWFs,
$a_{LR}=4.6a$ for WFs. Hence, the amount of required real-space volume decreases by a factor of $1.9$ to represent each wave function.

\begin{table}[t]
 \centering
 \caption{$\mathrm{Si}$: Fitting parameters of $\epsilon^{\infty}(a_{LR})$ function in Eq.~(\ref{Eq_epsfit}) to the WF and NGWF data
 in Fig.~\ref{fig_Si_eps}. The values from Ref.~\onlinecite{ref_polWf} are for calculations using orthogonal Wannier functions.
 The value of $\alpha$ in Ref.~\onlinecite{ref_polWf} is multiplied by a factor of $2$ to account for the $a_{LR} \mapsto 2a_{LR}'$
 relation between localization region size $a_{LR}$ used in our work and $a_{LR}'$ used in Ref.~\onlinecite{ref_polWf}.\label{tab_Si_epsfit}}
 \begin{ruledtabular}
 \begin{tabular*}{\linewidth}{@{\extracolsep{\fill} } cccc}
  Calculation & $\alpha$ & $A$ & $B$ \\
  \hline
  WF                          & 1.77 & -11.9 & 12.7 \\
  NGWF                        & 1.39 & -11.0 & 12.8 \\
  Ref.~\onlinecite{ref_polWf} & 1.62 & -13.2 & 13.4 \\
 \end{tabular*}
 \end{ruledtabular}
\end{table}

\begin{figure}[b]
\centering
\begin{tikzpicture}
\node[anchor=south west,inner sep=0] (image) at (0,0) {\includegraphics[width=0.99\linewidth]{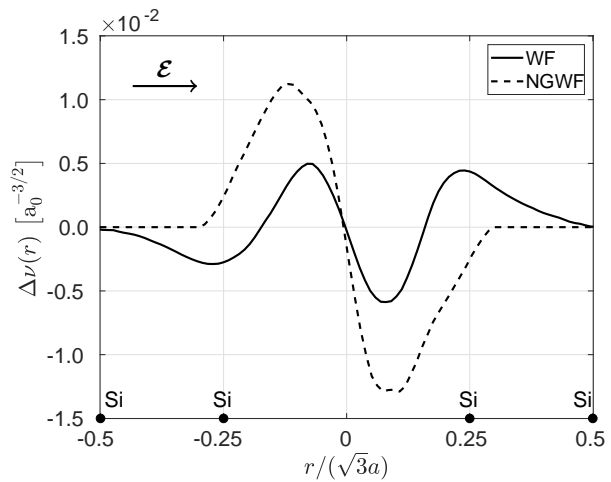}};
\begin{scope}[x={(image.south east)},y={(image.north west)}]
 \draw[thick,->] (0.19,0.82) -- (0.29,0.82) node[midway,above] {$\efield$};
\end{scope}
\end{tikzpicture}
\caption{$\mathrm{Si}$: Variation $\Delta{}\nu(r)=\nu(\efield;r)-\nu(r)$ of $\sigma$ wave functions
along \mbox{$\mathrm{Si}$--$\mathrm{Si}$} bond due to an electric field $\bar{e}|\efield|=10^{-2}\mathrm{\sfrac{Ry}{a_{0}}}$
applied in the direction parallel to the bond. Solid line: WF. Dashed line: NGWF.
The reference ground state wave functions $\nu(r)$ are displayed in Fig. \ref{fig_Si_wfcut}.
\label{fig_Si_dwfcut}}
\end{figure}

The significant impact of the localization constraint on the electronic response calculations reported above can be
better understood by examining of how the wave functions change under the action of an external electric field.
Fig.~\ref{fig_Si_dwfcut} shows the variation of the wave functions $\Delta{}\nu(r)=\nu(\efield;r)-\nu(r)$
along \mbox{$\mathrm{Si}$--$\mathrm{Si}$} bond due to an electric field $\bar{e}|\efield|=10^{-2}\mathrm{\sfrac{Ry}{a_{0}}}$
applied along the [111] direction, parallel to the bond.
The field-induced wave functions $\Delta{}\nu(r)$ are calculated with respect to the ground state ones $\nu(r)$
displayed in Fig.~\ref{fig_Si_wfcut}. Looking at Figs.~\ref{fig_Si_wfcut} and \ref{fig_Si_dwfcut} it can be seen that
because $\Delta{}\nu(r) \times \nu(r) \gtrless0$ for $r\lessgtr0$, the field-polarized wave functions
$\nu(\efield;r)$ are amplified at $r<0$ and damped at $r>0$. Therefore, the centroids of charge of the orbitals are shifted
with respect to the zero-field case. The displacement occurs in the positive direction of the $r$ axis, opposite to the applied field.
From Eq.~(\ref{eq_Pel_expr}) it can be concluded that this gives rise to an induced electronic polarization
$\Delta{}\vec{P}_{el}(\efield)=\vec{P}_{el}(\efield)-\vec{P}_{el}(\vec{0})$, in the direction compatible with the field.
It is used to quantify the dielectric response of the material according to Eq.~(\ref{Eq_eps}).
The results in Fig.~\ref{fig_Si_dwfcut} show that the field-induced wave functions
are better constrained within the localization region for non-orthogonal Wannier functions than for orthogonal ones.
This lead to (\emph{i}) a lower impact of the localization constraints when studying the electronic response
with NGWFs and (\emph{ii}) a more reliable value of the dielectric constant calculated for small localization regions.

\begin{figure}[h]
\subfloat[ground state WF\label{fig_Si_wf_gs}]{%
  \includegraphics[width=0.22\textwidth,clip=true]{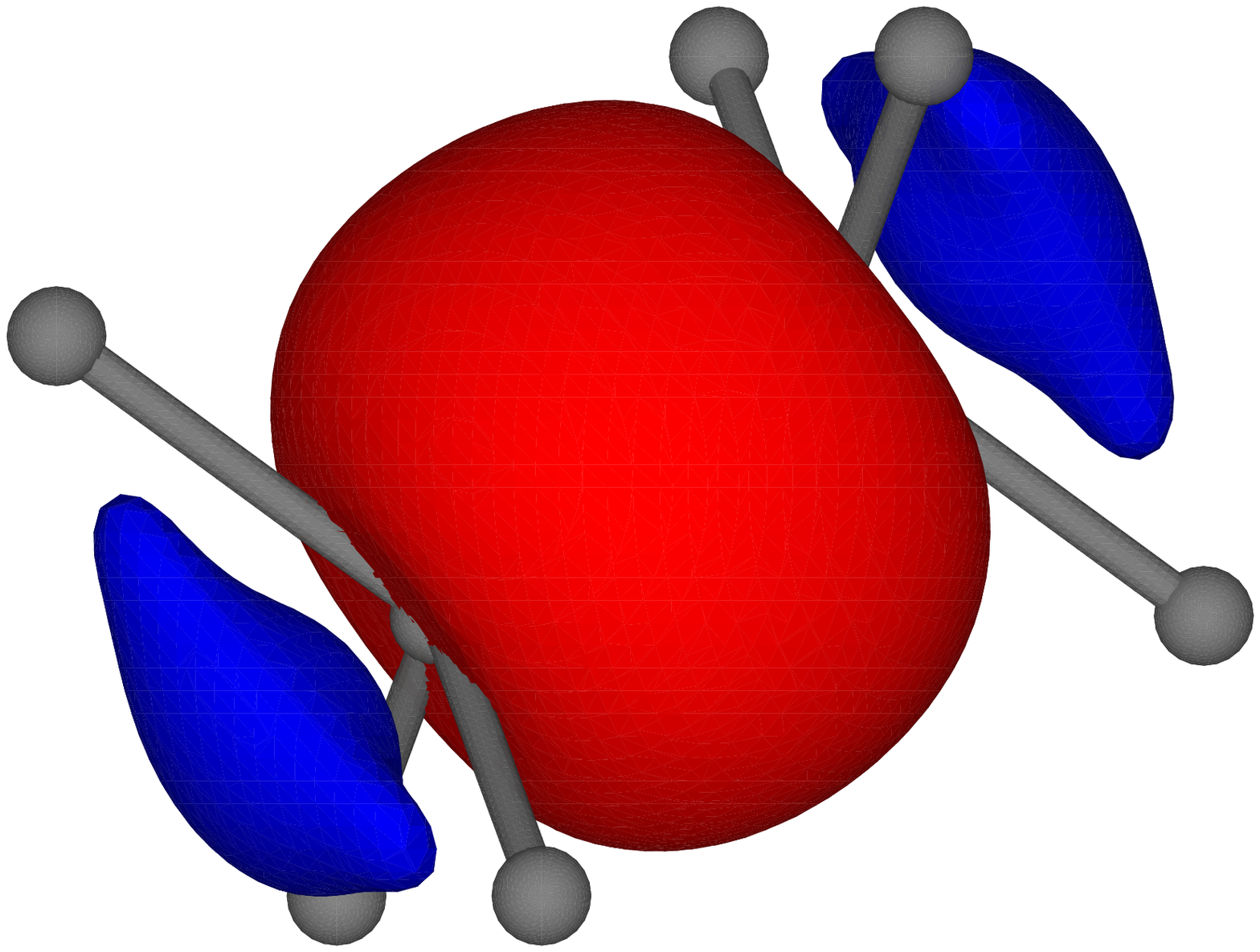}
}
\hfill
\subfloat[polarized WF\label{fig_Si_wf_field}]{%
  \begin{tikzpicture}
    \node[anchor=south west,inner sep=0] (image) at (0,0) {\includegraphics[width=0.22\textwidth,clip=true]{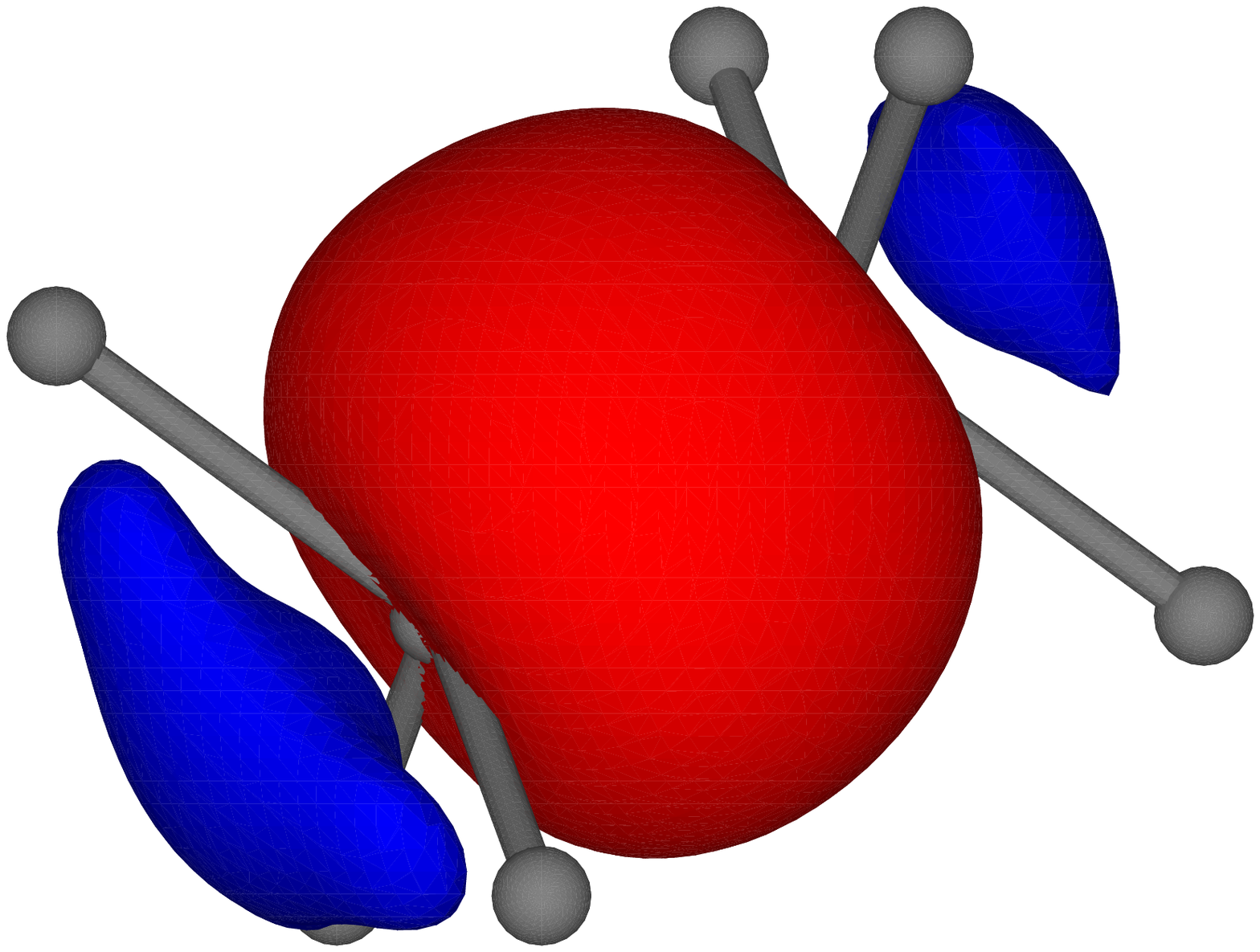}};
    \begin{scope}[x={(image.south east)},y={(image.north west)}]
      \draw[thick,->,>=stealth] (0.23,0.81) -- (0.37,0.92) node[above,left=2pt] {\small $\efield$};
    \end{scope}
  \end{tikzpicture}
}
\\
\smallskip
\subfloat[ground state NGWF\label{fig_Si_ngwf_gs}]{%
  \includegraphics[width=0.22\textwidth,clip=true]{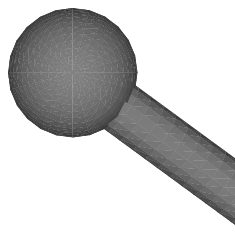}
}
\hfill
\subfloat[polarized NGWF\label{fig_Si_ngwf_field}]{%
  \begin{tikzpicture}
    \node[anchor=south west,inner sep=0] (image) at (0,0) {\includegraphics[width=0.22\textwidth,clip=true]{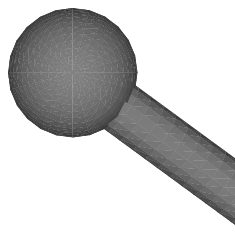}};
    \begin{scope}[x={(image.south east)},y={(image.north west)}]
      \draw[thick,->,>=stealth] (0.23,0.81) -- (0.37,0.92) node[above,left=2pt] {\small $\efield$};
    \end{scope}
  \end{tikzpicture}
}
\colorcaption{$\mathrm{Si}$: Amplitude isosurface plots of bond-centered $\sigma$ wave functions
at zero electric field (ground state orbitals) and in presence of macroscopic electric field
(polarized orbitals). The isosurfaces are taken at $\pm2\times10^{-3}\mathrm{a_{0}^{-3/2}}$ (red and blue
surfaces correspond to positive and negative amplitudes, respectively).
The two bonded $\mathrm{Si}$ atoms are shown (partially covered by the bonding wave function)
as well as their three remaining nearest-neighbors.
The polarized orbitals are induced by the electric field $\efield$ along [111] direction as indicated by arrows.
Drawings created with \MakeUppercase{VESTA} program\cite{ref_Vesta}.
\label{fig_Si_wf}}
\end{figure}

The ground state and field polarized Wannier functions of $\mathrm{Si}$ are shown in Fig.~\ref{fig_Si_wf}.
These $\sigma$ type orbitals are oriented along \mbox{$\mathrm{Si}$--$\mathrm{Si}$} bond. The orbitals are constrained
to be zero outside the LRs with size $a_{LR}=2a$. The polarized orbitals are inducted by an electric field
$\bar{e}|\efield|=10^{-2}\mathrm{\sfrac{Ry}{a_{0}}}$ applied along the bond in [111] direction.
It is encouraging to notice the similarity of the ground state WF in Fig.~\ref{fig_Si_wf_gs} to
maximally localized Wannier functions in bulk silicon\cite{ref_WFrev}. As is apparent in Fig.~\ref{fig_Si_wf_gs}
the ground state WF orbital represents the intuitive chemical concept of a covalent bond. It clearly displays
the character of the $\sigma$-bonding wave function created by the constructive interference of two $sp^3$
hybrid atomic orbitals centered on the bonded atoms. In addition, it can be seen in Fig.~\ref{fig_Si_wf_field}
that this covalent bond and its centroid of charge are shifted in the direction anti-parallel to the applied
electric field. The main difference in NGWFs as compared to WFs noticeable in Fig.~\ref{fig_Si_wf} is the absence
of $p$-like contributions on $\mathrm{Si}$, what makes NGWFs more localized and therefore better suited for
practical calculations. The better localization of NGWFs as compared to WFs is more apparent in Fig.~\ref{fig_Si_wfcut}
which shows the line cuts along rotation symmetry axis of the profiles displayed in  Fig.~\ref{fig_Si_wf_gs} (WF)
and Fig.~\ref{fig_Si_ngwf_gs} (NGWF).

\subsection{Cubic Barium Titanate} \label{sec_BTO}

\begin{figure}[b]
\centering
\includegraphics[width=0.99\linewidth]{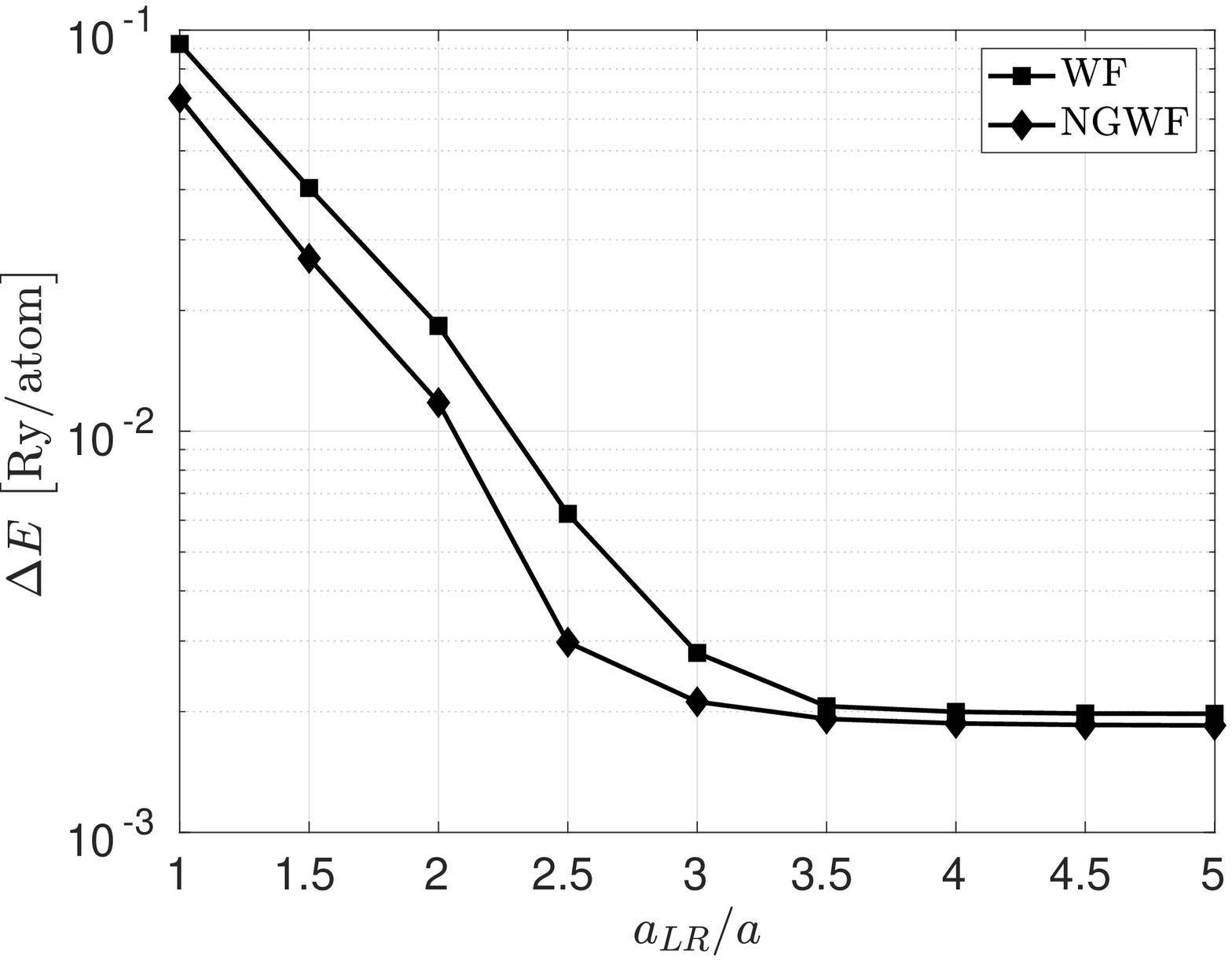}
\caption{$\mathrm{BaTiO_{3}}$: Error in band-structure energy $\Delta{}E_{bs}$ versus the size $a_{LR}$
of the localization regions. Squares: Wannier functions (WFs). Diamonds: non-orthogonal generalized
Wannier functions (NGWFs). $a_{LR}$ is normalized with the lattice constant $a=7.7\mathrm{a_{0}}$.}
\label{fig_BTO_dEbs}
\end{figure}

We now turn to a more complex system --- $\mathrm{BaTiO_{3}}$ in the centrosymmetric phase.
The simulation structure is a cubic perovskite unit cell. It is composed of
a $\mathrm{Ba}$ atom placed in the cube corner, a $\mathrm{Ti}$ atom sitting at the body-center
position, and $3$ $\mathrm{O}$ atoms occupying the face-centers of the perpendicular sides.
The lattice parameter used is $a=7.7\mathrm{a_{0}}$ ($\mathrm{a_{0}}$ denotes atomic length unit).
The $N=24$ valence electrons are covered by $12$ doubly occupied orbitals.
We consider localization regions of these orbitals centered on the $\mathrm{O}$ atoms and assign
$4$ localization centers to each of the $3$ $\mathrm{O}$ atoms. The $4$ orbitals in the overlaying
localization regions are initialized with Gaussians having a \emph{s}, \emph{px}, \emph{py} and \emph{pz}
symmetry, and origin on the central $\mathrm{O}$ atom. A grid spacing of $h=0.3\mathrm{a_{0}}$ and finite
difference discretization order $M=6$ is used. The chemical potential is set to $\mu=6\mathrm{Ry}$.
The reference energy calculations are performed using a $3\times3\times3$ Monkhorst-Pack mesh\cite{ref_MPgrid}
for the Brillouin Zone sampling.

Fig.~\ref{fig_BTO_dEbs} shows the convergence of the band-structure energy of $\mathrm{BaTiO_{3}}$
with respect to the size of localization $a_{LR}$. We observe that, similarly to the bulk silicon
case (see Fig.~\ref{fig_Si_dEbs}), allowing the wave functions to be non-orthogonal reduces the error
in the calculation due to the localization constraint. We note however a faster convergence in the case of
$\mathrm{BaTiO_{3}}$ than for $\mathrm{Si}$.

\begin{figure}[h]
\centering
\includegraphics[width=0.99\linewidth]{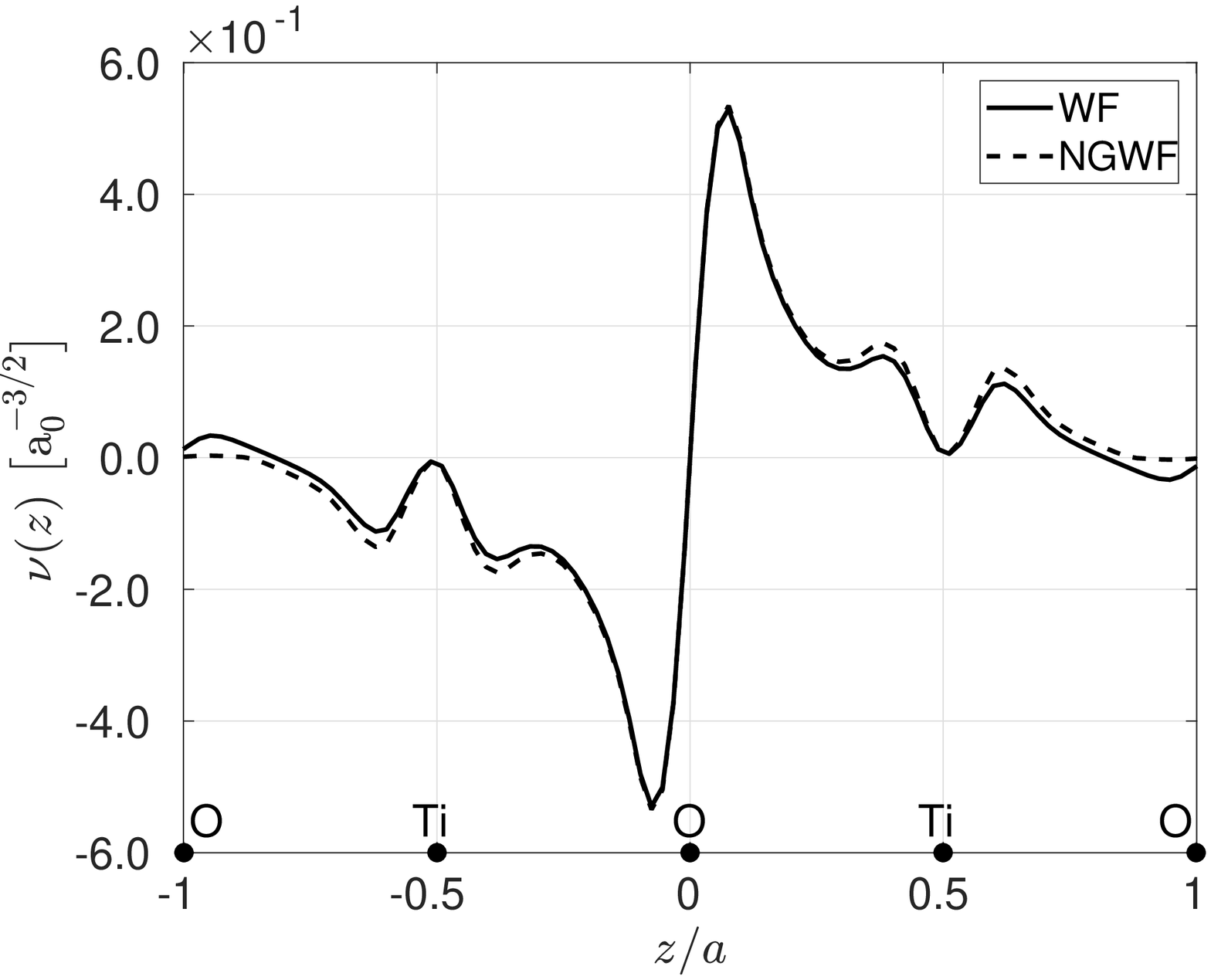}
\caption{$\mathrm{BaTiO_{3}}$: Line plot of Oxygen-centered $\sigma$ wave functions in ground state.
The orbitals are oriented parallel to the \mbox{$\mathrm{Ti}$--$\mathrm{O}$--$\mathrm{Ti}$} bond in the [001] direction.
$|z|$ is the distance from the localization center along the [001] axis. Solid line: WF. Dashed line: NGWF.
The wave functions are constrained to be zero outside localization region of size $a_{LR}=3a$.}
\label{fig_BTO_wfcut}
\end{figure}

The results presented in Fig.~\ref{fig_BTO_dEbs} are supported by carefully examining the profiles of the orbitals.
An exemplary line cut of the $\sigma$-type (\emph{pz}-initialized) orbital along a \mbox{$\mathrm{Ti}$--$\mathrm{O}$--$\mathrm{Ti}$}
bond in [001] direction is displayed in Fig.~\ref{fig_BTO_wfcut}. As it can be seen, the non-orthogonal wave function
is well contained within a distance $|z|<2a$. In contrary, the orthogonal one presents a significant amplitude around distant
$\mathrm{O}$ atoms located at $z=\pm 2a$. A similar behavior was observed for the other types of the orbitals.
As a consequence a larger LR is required for orthogonal wave functions than for non-orthogonal ones
in order to reduce the impact of the localization constraint on the quality of the calculations.

\begin{figure}[t]
\centering
\includegraphics[width=0.99\linewidth]{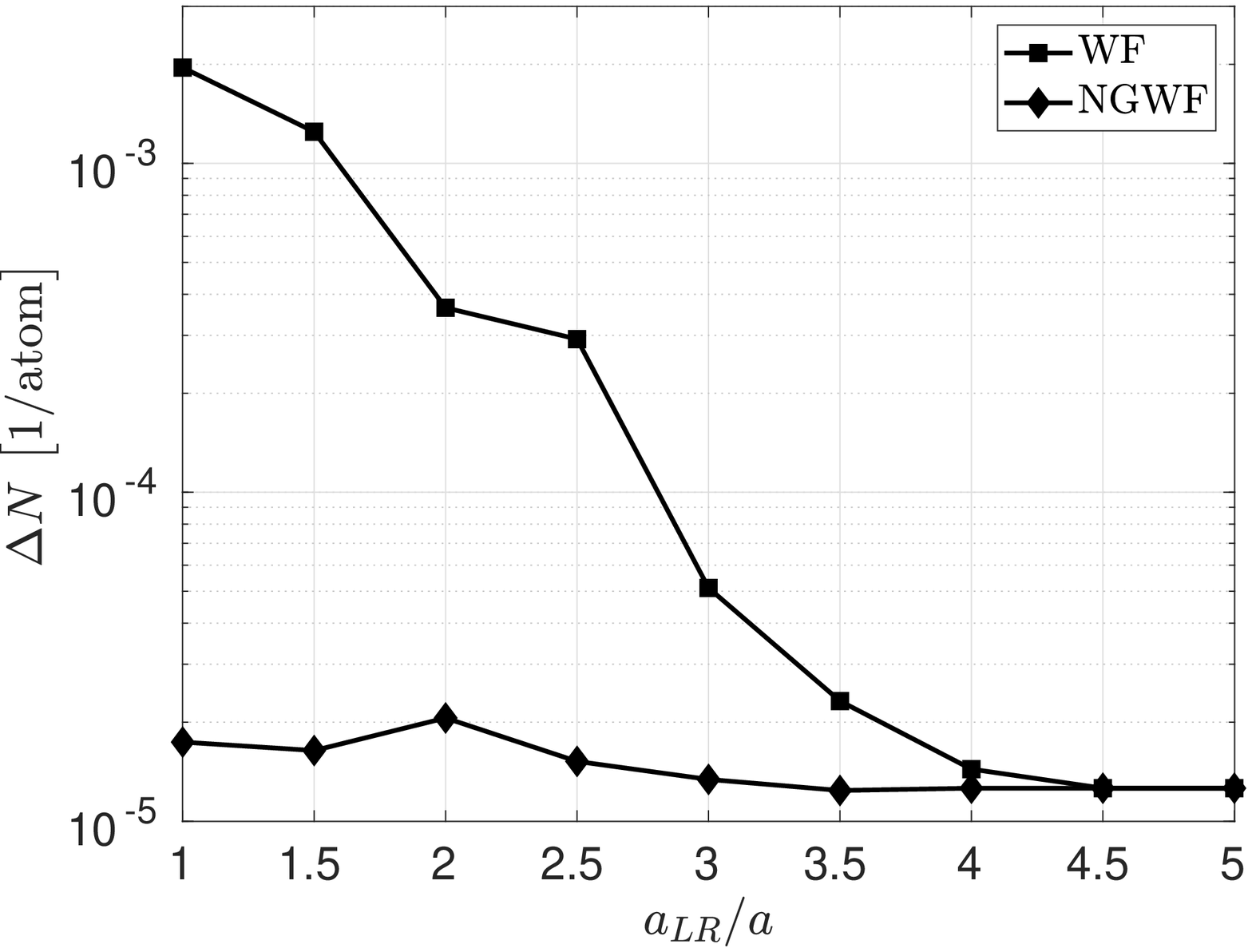}
\caption{$\mathrm{BaTiO_{3}}$: Error in total particle number $\Delta{}N$ versus the localization size $a_{LR}$.
Squares: WFs. Diamonds: NGWFs.}
\label{fig_BTO_dNtot}
\end{figure}

The error due to incompatible localization and orthogonality constraints manifests itself in the deviation of the total
particle number from its nominal valence value expressed by Eq.~(\ref{Eq_dNtot}). This can be observed in Fig.~\ref{fig_BTO_dNtot}.
As is apparent, in the limit of strong localization, the accuracy losses are severe when using WFs, but not for NGWFs for which
the inclusion of the $\mathbf{K}$ matrix compensates for the non-orthogonality of the wave functions.

\begin{figure}[b]
\centering
\includegraphics[width=0.99\linewidth]{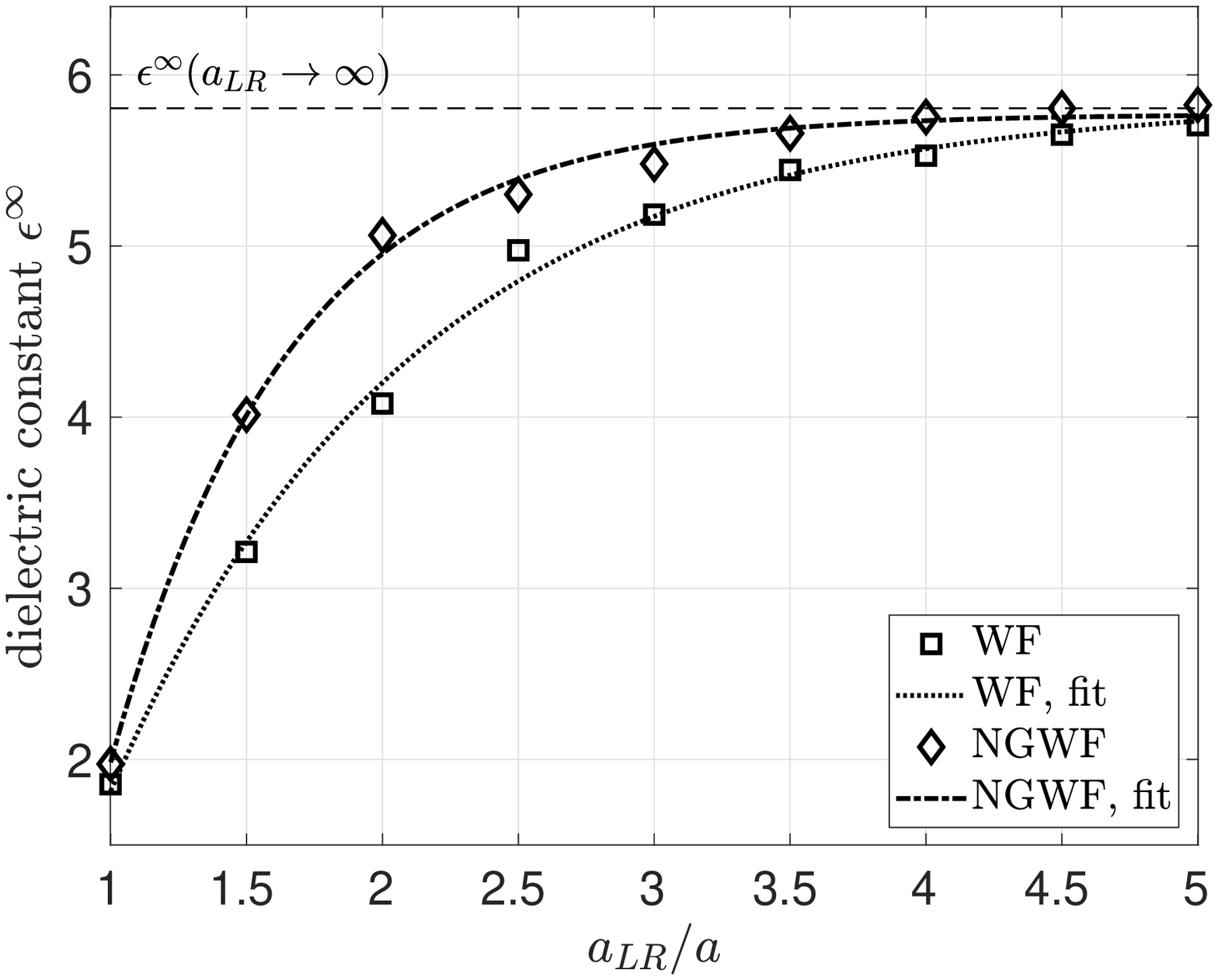}
\caption{$\mathrm{BaTiO_{3}}$: Electronic dielectric constant $\epsilon^{\infty}$ computed with with WFs (squares) and NGWFs (diamonds)
for different sizes of localization regions $a_{LR}$, normalized by the lattice constant $a=7.7\mathrm{a_{0}}$.
The dotted and dash-dot lines display fitting of Eq.~(\ref{Eq_epsfit}) to WF and NGWF data respectively.
The vertical dashed line show the average of the extrapolated fitted curves in the limit of $a_{LR}\rightarrow{}\infty$.}
\label{fig_BTO_eps}
\end{figure}

Fig.~\ref{fig_BTO_eps} presents the electronic dielectric constant of cubic $\mathrm{BaTiO_{3}}$ calculated using WFs and NGWFs
as a function of localization region size $a_{LR}$. Due to the inversion symmetry of the considered cubic perovskite structure it holds:
$\epsilon=\epsilon_{xx}=\epsilon_{yy}=\epsilon_{zz}$. In the calculations, the electric field is applied along the [001] direction,
by varying $\mathcal{E}_{z}$ coefficient, and the computed values of the induced $P_{el,z}$ polarization component are used to
evaluate the $\epsilon^{\infty}_{zz}$ tensor element according to Eq.~(\ref{Eq_eps}). The maximum intensity of the field is
$\bar{e}|\efield|=10^{-2}\mathrm{\sfrac{Ry}{a_0}}$. As can be seen in Fig.~\ref{fig_BTO_eps} the computed values of $\epsilon^{\infty}$
converge exponentially with increasing $a_{LR}$. Fitting the data using the function in Eq.~(\ref{Eq_epsfit}) gives in the limit
$a_{LR}\rightarrow{}\infty$ the values of $\epsilon^{\infty}$ equal to $5.77$ for WFs and $5.84$ for NGWFs.
Published LRT results vary between $5.60$ and $6.80$\cite{[{See }][{, and references therein.}]Ref_epsBTO_LRT}.
The experimental value is $5.4$\cite{Ref_epsBTO_exp}.

As can be seen in Fig.~\ref{fig_BTO_eps} the electronic response of $\mathrm{BaTiO_{3}}$ is practically converged at $a_{LR}$
equal to $3.8a$ and $2.7a$ when using WFs and NGWFs, respectively. In this case the error in $\epsilon^{\infty}$ is less than
$5\%$ of the average extrapolated value $\epsilon^{\infty}$ at $a_{LR}\rightarrow{}\infty$. This is quite different from the case of
$\mathrm{Si}$, which requires LRs that contain larger number of unit cells $a_{LR}/a$ to perform the calculations of the same quality.
As for $\mathrm{Si}$, by using non-orthogonal orbitals the impact of the localization constraint on the dielectric response is reduced.
In the case of $\mathrm{BaTiO_{3}}$ the volume fraction of LRs giving a relative error in $\epsilon^{\infty}$ of $5\%$
for WFs and NGWFs is $2.6$.

\begin{table}[h]
 \centering
 \caption{$\mathrm{BaTiO_{3}}$: Orbital decomposition of the induced electronic polarization $\mathrm{d}P_{el}=P_{el}(\efield)-P_{el}(\vec{0})$
due to electric field $\bar{e}|\efield|=10^{-2}\mathrm{\sfrac{Ry}{a_{0}}}$ along [001] direction for WFs and NGWFs.
These values correspond to relative contributions $\frac{\mathrm{d}P_{el,i}}{\mathrm{d}P_{el}}$ of the individual orbitals $i=1,\dots,12$.
The equivalent orbitals on the two $\mathrm{O}_{XY}$ atoms give the same contributions~to~$\mathrm{d}P_{el}$.}
 \begin{ruledtabular}
 \begin{tabular*}{\linewidth}{@{\extracolsep{\fill} } cccc}
  & & \multicolumn{2}{c}{Calculation} \\
  \cline{3-4}
  Atom & Orbital & WF & NGWF \\
  \hline
  $\mathrm{O}_{Z}$              &  $s$  & 0.003 & 0.002 \\
  $\mathrm{O}_{Z}$              & $p_x$ & 0.161 & 0.161 \\
  $\mathrm{O}_{Z}$              & $p_y$ & 0.161 & 0.161 \\
  $\mathrm{O}_{Z}$              & $p_z$ & 0.343 & 0.344 \\
  $\mathrm{O}_{XY} ~ (\times2)$ &  $s$  & 0.000 & 0.000 \\
  $\mathrm{O}_{XY} ~ (\times2)$ & $p_x$ & 0.041 & 0.040 \\
  $\mathrm{O}_{XY} ~ (\times2)$ & $p_y$ & 0.022 & 0.022 \\
  $\mathrm{O}_{XY} ~ (\times2)$ & $p_z$ & 0.103 & 0.104 \\
 \end{tabular*}
 \end{ruledtabular}
\label{tab_BTO_decomp}
\end{table}

To shed light on the charge transfer in $\mathrm{BaTiO_{3}}$ due to external electric field, we have listed in Table~\ref{tab_BTO_decomp}
a decomposition of the induced polarization coming from the individual orbitals. The orbitals are labeled by their dominant atomic character
on the $\mathrm{O}$ atom at the localization center. The $\mathrm{O}$ atoms are classified in two groups: the $\mathrm{O}$ atom along the [001] axis,
denoted as $\mathrm{O}_{Z}$, and the $\mathrm{O}$ atoms in (001) plane, named $\mathrm{O}_{XY}$.
The results listed in table~\ref{tab_BTO_decomp} show that the contributions from the same types of orbitals are similar for WFs and NGWFs.
We also note that the form of the decomposition is almost independent from the size of the localization region --- the changes in the relative
orbitals contributions are less than $1\%$ for $a_{LR}=[1,5]a$. As it stands out from a further inspection of the table, the $\mathrm{O}_{Z}(p_{z})$
orbital gives dominant contribution to the induced polarization, for the electric field applied in [001] direction. This orbital corresponds to the
$\sigma$ wave functions displayed in Fig.~\ref{fig_BTO_wfcut}.

\begin{figure}[t]
\centering
\begin{tikzpicture}
\node[anchor=south west,inner sep=0] (image) at (0,0) {\includegraphics[width=0.99\linewidth]{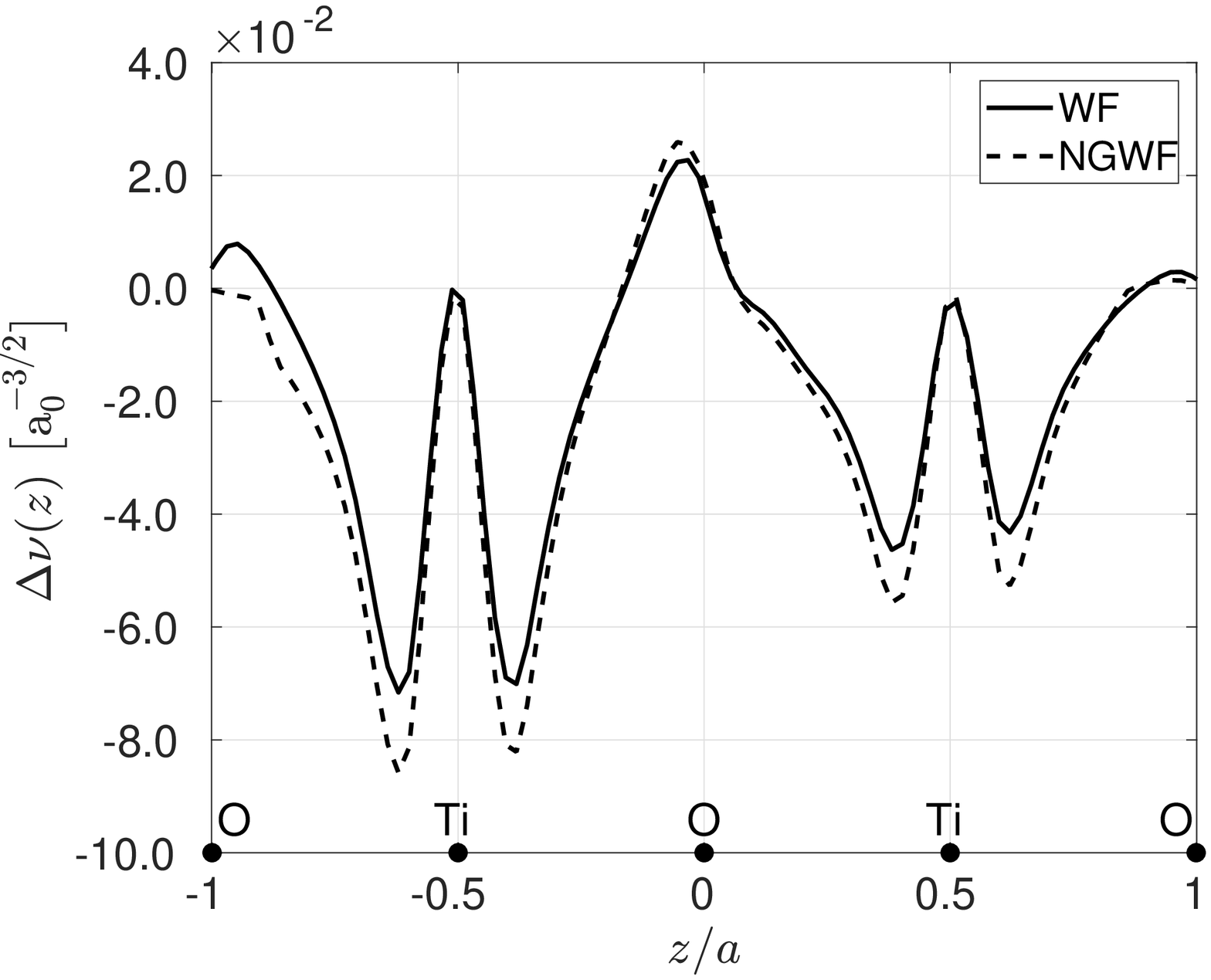}};
\begin{scope}[x={(image.south east)},y={(image.north west)}]
 \draw[thick,->] (0.20,0.83) -- (0.30,0.83) node[midway,above] {$\efield$};
\end{scope}
\end{tikzpicture}
\caption{$\mathrm{BaTiO_{3}}$: Variation $\Delta{}\nu(z)=\nu(\efield;z)-\nu(z)$ of $\sigma$ wave functions
along \mbox{$\mathrm{Ti}$--$\mathrm{O}$--$\mathrm{Ti}$} bond in the [001] direction due to an electric field
$\bar{e}|\efield|=10^{-2}\mathrm{\sfrac{Ry}{a_{0}}}$ applied in the direction parallel to the bond.
Solid line: WF. Dashed line: NGWF.
The corresponding ground state wave functions are displayed in Fig. \ref{fig_BTO_wfcut}.}
\label{fig_BTO_dwfcut}
\end{figure}

Fig.~\ref{fig_BTO_dwfcut} shows how the $\sigma$ wave functions of $\mathrm{BaTiO_{3}}$ change under an applied electric field.
As it can be seen, for the orthogonal wave function a large charge transfer is present around distant $\mathrm{O}$ atoms located at
$z=\pm 2a$. In consequence by using $a_{LR} < 2a$ leads to a significantly reduced electronic response and results in an underestimated
value of $\epsilon^{\infty}$, as reported in Fig.~\ref{fig_BTO_eps}. Because non-orthogonal wave functions are only slightly perturbed
by the electric field at distances $|z|>2a$, a better convergence of the dielectric constant calculations is obtained with NGWFs than by
using WFs.

\begin{figure}[b]
\subfloat[ground state WF\label{fig_BTO_wf_gs}]{%
  \includegraphics[width=0.22\textwidth,clip=true]{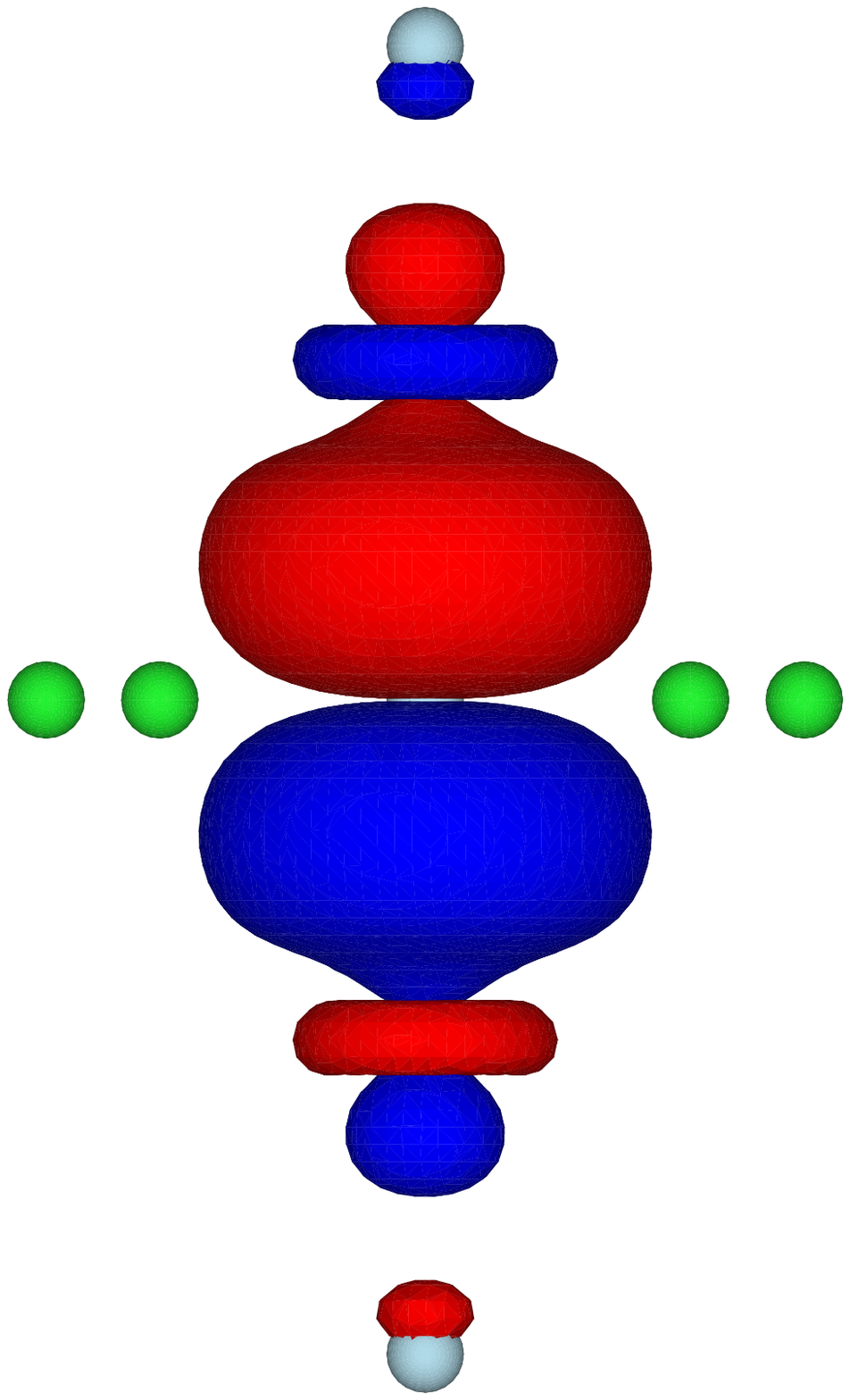}
}
\hfill
\subfloat[polarized WF\label{fig_BTO_wf_field}]{%
  \begin{tikzpicture}
    \node[anchor=south west,inner sep=0] (image) at (0,0) {\includegraphics[width=0.22\textwidth,clip=true]{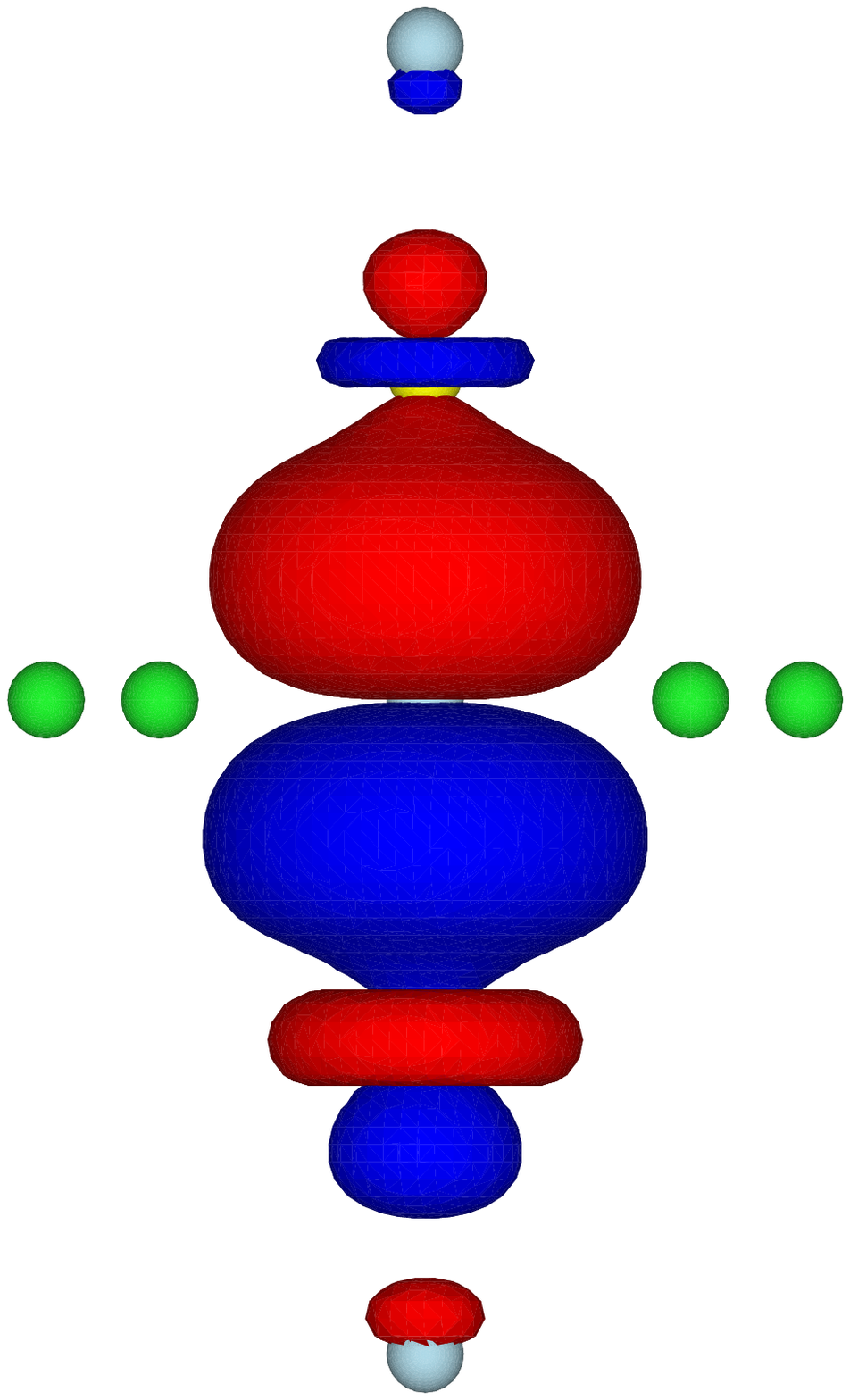}};
    \begin{scope}[x={(image.south east)},y={(image.north west)}]
      \draw[thick,->,>=stealth] (0.20,0.79) -- (0.20,0.92) node[midway,right] {\small $\efield$};
    \end{scope}
  \end{tikzpicture}
}
\\
\smallskip
\subfloat[ground state NGWF\label{fig_BTO_ngwf_gs}]{%
  \includegraphics[width=0.22\textwidth,clip=true]{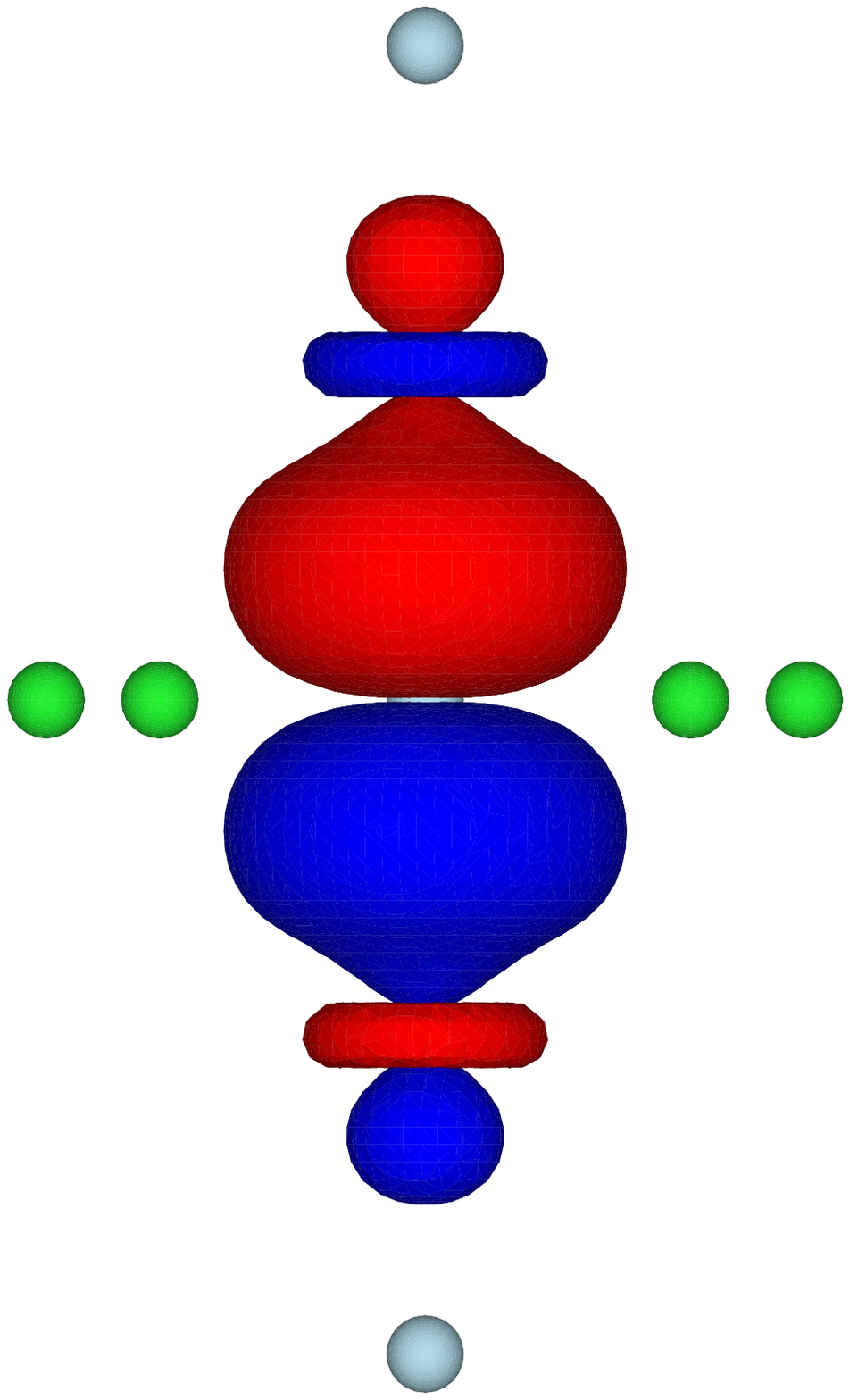}
}
\hfill
\subfloat[polarized NGWF\label{fig_BTO_ngwf_field}]{%
  \begin{tikzpicture}
    \node[anchor=south west,inner sep=0] (image) at (0,0) {\includegraphics[width=0.22\textwidth,clip=true]{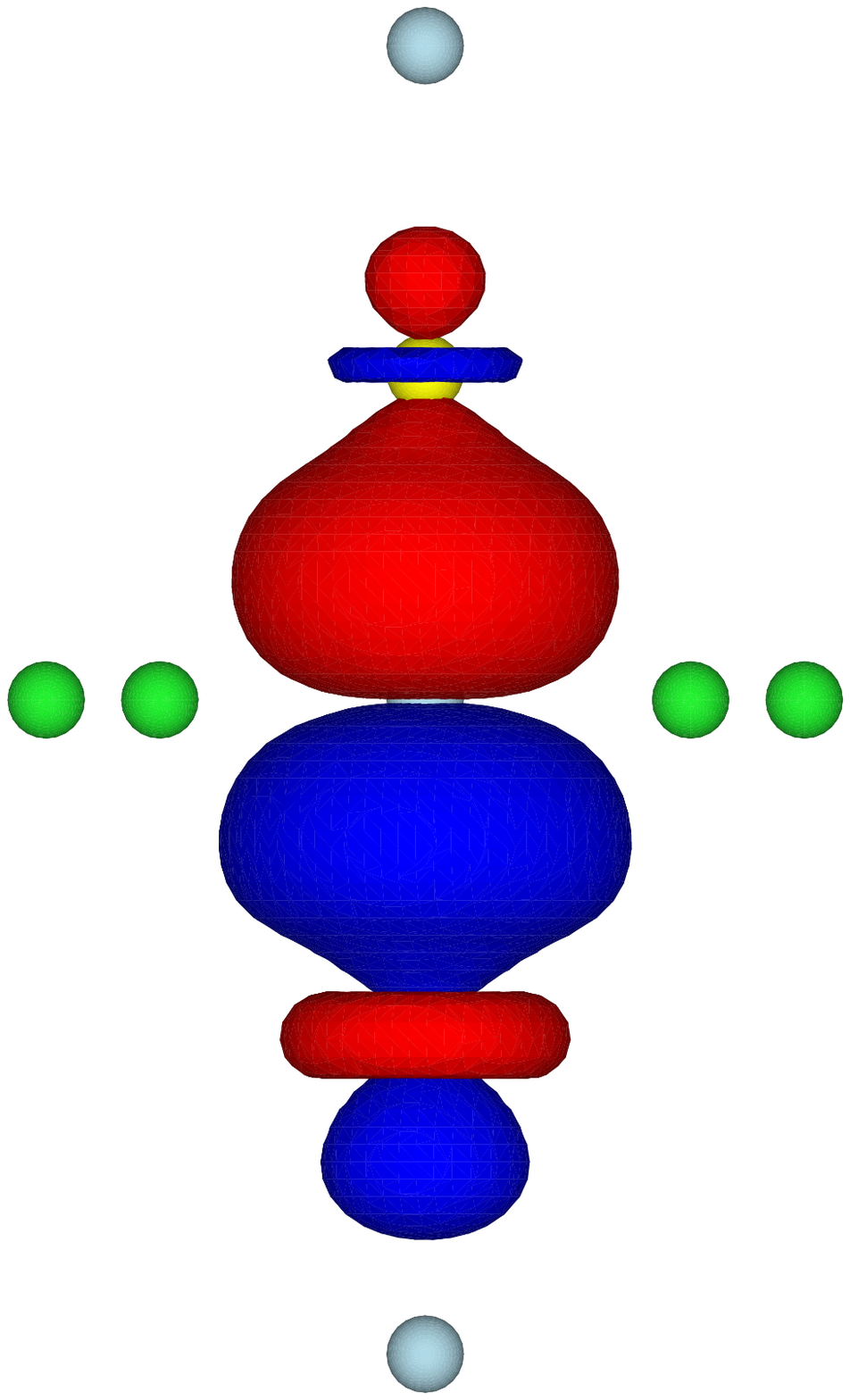}};
    \begin{scope}[x={(image.south east)},y={(image.north west)}]
      \draw[thick,->,>=stealth] (0.20,0.79) -- (0.20,0.92) node[midway,right] {\small $\efield$};
    \end{scope}
  \end{tikzpicture}
}
\colorcaption{$\mathrm{BaTiO_{3}}$: Amplitude isosurface plots of oxygen-centered $\sigma$ wave functions at zero electric field
(ground state orbitals) and in presence of macroscopic electric field (polarized orbitals).
Isosurfaces at $\pm2\times10^{-3}\mathrm{a_{0}^{-3/2}}$ (red and blue surfaces correspond to positive and negative amplitudes respectively).
The orbitals are oriented along $\mathrm{O}$--$\mathrm{Ti}$--$\mathrm{O}$--$\mathrm{Ti}$--$\mathrm{O}$ chains in the [001] direction.
A $\mathrm{O}$ atom is at the center, embedded in a $p_{z}$ orbital; above and below are $\mathrm{Ti}$ atoms (yellow),
almost hidden under $d_{z^2}$ orbitals; the two other $\mathrm{O}$ atoms (light blue) are visible on top and bottom.
The four $\mathrm{Ba}$ atoms (light green) neighboring the central oxygen are also shown.
The polarized orbitals are induced by the electric field $\efield$ along [001] direction as indicated by arrows.
Drawings created with \MakeUppercase{VESTA} program\cite{ref_Vesta}.}
\label{fig_BTO_wf}
\end{figure}

Finally, in Fig.~\ref{fig_BTO_wf} we show the isosurfaces of the $\sigma$ orbitals oriented along \mbox{$\mathrm{Ti}$--$\mathrm{O}$--$\mathrm{Ti}$} bond
in [001] direction. They result from $p_z$ atomic orbitals centered on $\mathrm{O}$ atom along [001] axis, after minimization of the electric enthalpy
functional under the localization constraint $a_{LR}=3a$. The ground state orbitals correspond to zero field calculation and the polarized orbitals
are induced by an electric field $\bar{e}|\efield|=10^{-2}\mathrm{\sfrac{Ry}{a_{0}}}$ applied in the [001] direction.
We note the similarity of the ground state WF in Fig.~\ref{fig_BTO_wf_gs} to the corresponding MLWF in centrosymmetric Barium Titanate\cite{ref_btoWf}.
As it can be seen in Fig.~\ref{fig_BTO_wf_gs}, the wave function show clearly the hybridization between $p_{z}$ orbital on the $\mathrm{O}$ atom in the center
and $d_{z^2}$ orbitals on the neighboring $\mathrm{Ti}$ atoms. The hybridization to $\mathrm{Ti}$ $d_{z^2}$ states appears in the form of
tori surrounding the $\mathrm{Ti}$ atoms (in Fig.~\ref{fig_BTO_wf} the $\mathrm{Ti}$ atoms are embedded in the $d_{z^2}$ orbitals).
Such hybridization is at the origin of the ferroelectric instability as argued by Posternak \emph{et~al.}\cite{ref_ferroOrigin}.

The application of the electric field changes the chemical bonding, as indicated in Fig.~\ref{fig_BTO_wf}. For the electric field
acting in the [001] direction, the hybridization weakens for the upper $\mathrm{O}$--$\mathrm{Ti}$ bond and strengthens
for the lower one, endowing the wave functions with less $d_{z^2}$ character on the top than on the bottom.
This feature is captured by both WFs and NGWFs. 

The main difference between WFs and NGWFs is the presence of the $p_{z}$-like contributions at distant $\mathrm{O}$ atoms on the top and bottom of the figures,
in the case of orthogonal Wannier functions. The better localization of NGWFs as compared to WFs is also apparent in Fig.~\ref{fig_Si_wfcut} which plots the
line cuts along rotation symmetry axis of the profiles displayed in Fig.~\ref{fig_BTO_wf_gs} (WF) and Fig.~\ref{fig_BTO_ngwf_gs} (NGWF). When the electric field
is applied in the [001] direction, a charge transfer occurs between these contributions, which can be seen by comparing Fig.~\ref{fig_BTO_wf_gs} and Fig.~\ref{fig_BTO_wf_field}.
As a consequence, a localization region containing at least $3$ unit cells is necessary to perform qualitatively accurate calculations with WFs. On the contrary,
for non-orthogonal Wannier functions the charge transfer occurs only in the main body of the wave function, contained within the utmost $\mathrm{O}$ atoms
(comparison of Figs.~\ref{fig_BTO_ngwf_gs} and \ref{fig_BTO_ngwf_field}). This alleviates the impact of the localization constraint on the accuracy of the
calculations employing NGWFs.

\section{Conclusions} \label{sec_conclusions}
We have developed a formalism for calculating the response of an insulator to a static, homogeneous
electric field based on an optimization of non-orthogonal generalized Wannier functions.
It extends the NV approach to finite electric fields in which orthogonal Wannier functions
are used to write a functional for the electric enthalpy of a solid in a uniform
electric field. We have implemented this formalism in a fully self-consistent pseudopotential
LDA scheme and applied it to representative systems. This has allowed us to asses the
practical usefulness of the method.

The analysis carried out has demonstrated the ability of polarized Wannier functions to highlight the changes
of chemical bonding in solids due to applied electric field. As has also been shown, the localized orbitals
allow for an intuitive understanding of the effects of the field in terms of displacements of centroids of charge
of the wave functions. Therefore a decomposition of the electronic response coming from the individual orbitals
is readily available. The main qualitative features are shared between orthogonal and non-orthogonal orbitals.
However our results have clearly demonstrated that the higher localization of non-orthogonal wave functions
does not affect the physical results.

As future developments, the proposed method could possibly be extended to use more orbitals
than the number of occupied bands. The density kernel matrix would then play the role of generalized occupation
numbers\cite{ref_WFrev}. The inclusion of extra orbitals would enable long-range charge transfers in
the minimization process, irrespective of the extent of the localization regions of the wave functions.
This has been shown to decrease the error in the variational estimate of the ground state energy in the
context of zero field calculations based on localized Bloch-like orbitals\cite{ref_KMG}. For the finite field
calculations the ability of working with truly localized Wannier-like non-orthogonal orbitals,
as in our formulation, is a necessary first step.

\begin{acknowledgments}
This work was supported by SNF Grants No. \texttt{PP00P2\_159314} and \texttt{200021\_149495}.
\end{acknowledgments}

\appendix
\section{Use of the Chemical Potential to Ensure the Variational Property of the\\Minimized Energy Functional\label{sec_app_chempot}}

In this Appendix we justify the approach consisting of shifting the eigenspectrum of the
Hamiltonian operator, to make its eigenvalues negative. This is done by using the chemical
potential parameter $\mu$ in the optimization procedure, as introduced in Sec.~\ref{sec_compdetails}.
To this end, the minimized energy functional is compared with the exact one and its
variational properties are revealed.

The single-particle density matrix written for overlapping orbitals is given by \cite{ref_largeScale_overlapping}
\begin{equation}
 \hat{\rho}[\{\nu\},\mathbf{S}^{-1}] = 2 \sum_{ab} \ket{\nu_a} (\mathbf{S}^{-1})_{ab} \bra{\nu_b} ~,
 \label{eq_densop_exact}
\end{equation}
where $\mathbf{S}^{-1}$ is the inverse of the overlap matrix $\mathbf{S}$ in Eq.~(\ref{eq_Smat}),
$\mathbf{S}^{-1}\times\mathbf{S}=\mathbf{I}$.
For notational simplicity the indexing over cell replicas is dropped in this Appendix.

The exact electronic energy $W[\{\nu\},\mathbf{S}^{-1}]$, evaluated as the trace of the product of the
density matrix in Eq.~(\ref{eq_densop_exact}) and the Hamiltonian operator in Eq.~(\ref{eq_Hfield}),
can be written~as
\begin{equation}
 W[\{\nu\},\mathbf{S}^{-1}] = 2 \sum_{ab} (\mathbf{S}^{-1})_{ab} \bra{\nu_a} \hat{H} \ket{\nu_b} ~.
 \label{eq_W_exact}
\end{equation}
Note that Eq.~(\ref{eq_W_exact}) corresponds to the DFT energy functional written for overlapping orbitals.
It is used in large scale electronic structure calculations \cite{ref_largeScale_overlapping}
and \emph{ab initio} molecular dynamics simulations \cite{ref_MD_overlapping}, under zero electric-field
conditions.

The expression for the electronic enthalpy introduced in Eq.~(\ref{eq_W_expr}) and used in the minimization procedure
of Sec.~\ref{sec_compdetails} is restated below
\begin{equation}
 W[\{\nu\},\mathbf{Q}] = 2\sum_{ab} Q_{ab} \bra{\nu_a} \hat{H} \ket{\nu_b} ~,
 \label{eq_W_approx}
\end{equation}
where
\begin{equation}
 \mathbf{Q}=2\mathbf{K}-\mathbf{K}\times\mathbf{S}\times\mathbf{K} ~,
 \label{eq_Qmat_approx}
\end{equation}
as given in Eq.~(\ref{eq_Qmat_def}).

This energy functional corresponds to the transformed density operator defined in Eq.~(\ref{eq_densop_transf})
\begin{equation}
 \hat{\rho}[\{\nu\},\mathbf{Q}] = 2 \sum_{ab} \ket{\nu_a} Q_{ab} \bra{\nu_b} ~.
 \label{eq_densop_approx}
\end{equation}

The difference between the approximate and exact energy functionals
\begin{equation}
 \Delta{}W = W[\{\nu\},\mathbf{Q}]-W[\{\nu\},\mathbf{S}^{-1}] ~,
 \label{eq_W_diff}
\end{equation}
can be calculated with the method of Mauri \emph{et al.} \cite{ref_MGC}.
In~this approach, the changes in the energy functional are parameterized with respect to a dimensionless
parameter~$\lambda$. The later varies continuously from zero, which corresponds to $W[\{\nu\},\mathbf{S}^{-1}]$,
to one, which is equivalent to $W[\{\nu\},\mathbf{Q}]$. Hence, Eq.~(\ref{eq_W_diff}) can be written as
\begin{equation}
 \Delta{}W = \int_0^1 \frac{\partial W[\{\nu\},\mathbf{A}(\lambda)]}{\partial \lambda} ~\mathrm{d}\lambda ~,
 \label{eq_W_diff_int}
\end{equation}
where $\mathbf{A}(\lambda)=\lambda\cdot\left(\mathbf{Q}-\mathbf{S}^{-1}\right) + \mathbf{S}^{-1}$.

By combining Eqs.~(\ref{eq_W_exact}) and (\ref{eq_W_approx}), Eq.~(\ref{eq_W_diff_int}) can be evaluated as
\begin{equation}
 \Delta{}W = \sum_{ab} (\mathbf{Q}-\mathbf{S}^{-1})_{ab} \bra{\nu_a} \hat{\bar{H}} \ket{\nu_b} ~,
 \label{eq_W_diff_average}
\end{equation}
where $\hat{\bar{H}}$ is the $\lambda$-averaged Hamiltonian operator, given~by
\begin{equation}
 \hat{\bar{H}} = -\frac{1}{2}\nabla^2 + \hat{V}_{ext} + \int_0^1 \hat{V}_{HXC}(\lambda) ~\mathrm{d}\lambda ~.
 \label{eq_Hamiltonian_average}
\end{equation}
Here, $\hat{V}_{ext}=\hat{V}_{ion}+\efield\cdot\vec{r}$ and
$\hat{V}_{HXC}(\lambda)=\hat{V}_{H}(\lambda)+\hat{V}_{XC}(\lambda)$,
where the Hartree and exchange-correlation potentials are calculated using the charge density
$\rho[\{\nu\},\mathbf{A}(\lambda)]$,
when integrating over~$\lambda$.

By recalling the definition of the $\mathbf{Q}$ matrix, repeated in Eq.~(\ref{eq_Qmat_approx}), the matrix $(\mathbf{Q}-\mathbf{S}^{-1})$
appearing in Eq.~(\ref{eq_W_diff_average}) can be expressed as
\begin{equation}
 (\mathbf{Q}-\mathbf{S}^{-1}) = -\mathbf{S}^{-1} \times \left( \mathbf{I} - \mathbf{S} \times \mathbf{K} \right)^2 ~.
 \label{eq_Qmat_approx_ND}
\end{equation}
The above form shows that $(\mathbf{Q}-\mathbf{S}^{-1})$ is negative-definite~(ND) --- it can be seen from Eq.~(\ref{eq_Qmat_approx_ND}) that
it is the negation of a product of two positive-definite~(PD) matrices. The $\mathbf{S}^{-1}$ matrix is PD since the overlap matrix~$\mathbf{S}$
possesses this property and every PD matrix is invertible and its inverse is also PD \cite{Ref_matrix_PD}. The PD property of the second term in
Eq.~(\ref{eq_Qmat_approx_ND}) follows directly from the fact that it is the square of a matrix. A similar line of argument can be used to prove
that the $(\mathbf{Q}-\mathbf{S}^{-1})$ matrix is ND too when optimizing the orthogonal Wannier functions with $\mathbf{K}=\mathbf{I}$.

Given a finite basis set, one can choose the $\mu$ parameter large enough so that
all eigenvalues of the operator
\begin{equation}
 \hat{\bar{H}}(\mu) = \hat{\bar{H}} - \hat{I}\mu
 \label{eq_Hop_mu}
\end{equation}
are negative, $\hat{\bar{H}}(\mu)\prec{}0$. Then, the $(N/2\times{}N/2)$ matrix
$\bra{\nu_a}(\hat{\bar{H}}-\hat{I}\mu)\ket{\nu_b}$
is also ND.
This requirement defines the chemical potential parameter $\mu$.

By substituting Eq.~(\ref{eq_Hop_mu}) into Eq.~(\ref{eq_W_diff_average}) it can be verified
that the eigenspectrum shift of the Hamiltonian ensures that $\Delta{}W$ is non-negative,
because it is equal to the trace of a product of two ND matrices.
This proves that if $\mu$ satisfies $\hat{\bar{H}}(\mu)\prec{}0$, then it holds:
\begin{equation}
 W[\{\nu\},\mathbf{Q}] \geq W[\{\nu\},\mathbf{S}^{-1}] ~.
 \label{eq_energy_geq}
\end{equation}
The above inequality gives the desired variational property. It ensures that our variational principle
in Eq.~(\ref{eq_W_approx}) has the exact Kohn-Sham ground-state energy as its absolute minimum.
Consequently, no spurious solutions are generated.

The equality in Eq.~(\ref{eq_energy_geq}) holds for each set of $\{\nu\}$, when $\mathbf{K}=\mathbf{S}^{-1}$,
as can be seen from Eqs.~(\ref{eq_W_exact}), (\ref{eq_W_approx}), and (\ref{eq_Qmat_approx}).
In this case $\Delta{}W=0$, as is apparent from Eqs.~(\ref{eq_W_diff_average}) and (\ref{eq_Qmat_approx_ND}).
Thus, the auxiliary matrix $\mathbf{K}$ at the minimum becomes a generalized inverse of the overlap matrix of the localized orbitals.
The property \mbox{$\mathbf{K}=\mathbf{S}^{-1}$} results in a weakly idempotent density matrix,
$\hat{\rho}^2=\hat{\rho}$, as can be concluded from Eqs.~(\ref{eq_Qmat_approx}) and (\ref{eq_densop_approx}).

Finally, we note that when the functional $W[\{\nu\},\mathbf{Q}]$ is minimized with respect to $\{\nu\}$,
with the density kernel fixed and set to the identity matrix, i.e. $\mathbf{K}\equiv\mathbf{I}$,
the equality in Eq.~(\ref{eq_energy_geq}) can be realized by $\mathbf{S}=\mathbf{I}$. This follows from
Eqs.~(\ref{eq_W_exact}), (\ref{eq_W_approx}), and (\ref{eq_Qmat_approx}), with $\mathbf{Q}=2\mathbf{I}-\mathbf{S}$.
Thus, the optimized orbitals are orthogonal. In this case, the functional in Eq.~(\ref{eq_W_approx})
corresponds to the one derived by Mauri~\emph{et~al.} \cite{ref_MGC} and Ordejon \emph{et~al.} \cite{ref_Ordejon_unconstrained}.
In our approach, by varying $\mathbf{K}$, the optimized orbitals are allowed to be non-orthogonal, 
which improves their localization.

\section{Position Operator in\\Extended Systems\label{sec_app_posop}}

The position operator in extended systems has been investigated in detail by Resta in Ref.~\onlinecite{ref_illPosedPosOp}.
In this Appendix we summarize the main results relevant in the context of finite-field calculations.

Within the Schr\"{o}dinger representation the result of the position operator, $\hat{\vec{r}}$, acting on a
wave function, $\phi$, equals the coordinate function, $\vec{r}$, multiplied by the wave function,
$\big(\hat{\vec{r}}\phi\big)(\vec{r})=\vec{r}\phi(\vec{r})$. \cite{ref_QMintro}
This applies only to localized orbitals which belong to the class of square-integrable wave functions.
In the basis of periodic Bloch functions, $\{\psi\}$, this operation becomes ill-defined because of the
following argument. The Hilbert space of the single particle wave functions is determined by the condition
$\psi(\vec{r}+\vec{R})=\psi(\vec{r})$, where the lattice vector $\vec{R}$ specifies the imposed periodicity.
An~operator maps any function of the given space into another function belonging to the same space.
This cannot be true for the position operator acting on a Bloch wave~function, $\psi$, since
\begin{equation}
 \vec{r} \psi(\vec{r}) \neq \big( \vec{r} + \vec{R} \big) \psi(\vec{r}+\vec{R}) ~.
 \label{eq_pos_Bloch}
\end{equation}
As can be seen from Eq.~(\ref{eq_pos_Bloch}) the multiplicative position operator $\hat{\vec{r}}$
is not a legitimate operator when periodic boundary conditions are adopted for the Bloch functions,
since $\hat{\vec{r}} \psi(\vec{r})$ is not a periodic function, even if $\psi(\vec{r})$ is.

This problem was addressed by Resta \cite{ref_illPosedPosOp} who proposed to define the expectation value
of the position operator in periodic systems by using the Berry phase approach, with much of the conceptual
work stemming from the earlier development of the modern theory of polarization \cite{ref_MTP}.
One of the most relevant features of this method is that the position operator in an extended quantum system
within periodic boundary conditions is no longer a single-particle operator: it acts as a genuine
many-body operator on the periodic wave function of $N$ electrons.
This~renders its implementation particularly challenging.

On the contrary, the position operator can be readily evaluated in the basis of Wannier-like functions.
The matrix elements of the position operator in this representation can be calculated directly,
using real-space integrals
\begin{equation}
 \bra{\nu_a^i} \hat{\vec{r}} \ket{\nu_b^j} = \int \nu_a^0(\vec{r}+\vec{R}_i) ~\nu_b^0(\vec{r}+\vec{R}_j) ~\vec{r} ~\mathrm{d}V ~,
 \label{eq_posint_Wannier}
\end{equation}
where the periodicity relation in Eq.~(\ref{eq_wf_period}) is used to express the remaining
orbitals in terms of the electronic degrees of freedom, which are indicated by the superscript~$0$
and~centered in the unit cell containing the origin. In practical calculations the
integration takes place over the part of space where the two localized orbitals overlap.
Since the wave functions are truncated to finite localization regions this operation is
well-defined.

\bibliography{ngwf_dielectric.bib}

\providecommand{\noopsort}[1]{}\providecommand{\singleletter}[1]{#1}%
\begin{thebibliography}{57}%
\makeatletter
\providecommand \@ifxundefined [1]{%
 \@ifx{#1\undefined}
}%
\providecommand \@ifnum [1]{%
 \ifnum #1\expandafter \@firstoftwo
 \else \expandafter \@secondoftwo
 \fi
}%
\providecommand \@ifx [1]{%
 \ifx #1\expandafter \@firstoftwo
 \else \expandafter \@secondoftwo
 \fi
}%
\providecommand \natexlab [1]{#1}%
\providecommand \enquote  [1]{``#1''}%
\providecommand \bibnamefont  [1]{#1}%
\providecommand \bibfnamefont [1]{#1}%
\providecommand \citenamefont [1]{#1}%
\providecommand \href@noop [0]{\@secondoftwo}%
\providecommand \href [0]{\begingroup \@sanitize@url \@href}%
\providecommand \@href[1]{\@@startlink{#1}\@@href}%
\providecommand \@@href[1]{\endgroup#1\@@endlink}%
\providecommand \@sanitize@url [0]{\catcode `\\12\catcode `\$12\catcode
  `\&12\catcode `\#12\catcode `\^12\catcode `\_12\catcode `\%12\relax}%
\providecommand \@@startlink[1]{}%
\providecommand \@@endlink[0]{}%
\providecommand \url  [0]{\begingroup\@sanitize@url \@url }%
\providecommand \@url [1]{\endgroup\@href {#1}{\urlprefix }}%
\providecommand \urlprefix  [0]{URL }%
\providecommand \Eprint [0]{\href }%
\providecommand \doibase [0]{http://dx.doi.org/}%
\providecommand \selectlanguage [0]{\@gobble}%
\providecommand \bibinfo  [0]{\@secondoftwo}%
\providecommand \bibfield  [0]{\@secondoftwo}%
\providecommand \translation [1]{[#1]}%
\providecommand \BibitemOpen [0]{}%
\providecommand \bibitemStop [0]{}%
\providecommand \bibitemNoStop [0]{.\EOS\space}%
\providecommand \EOS [0]{\spacefactor3000\relax}%
\providecommand \BibitemShut  [1]{\csname bibitem#1\endcsname}%
\let\auto@bib@innerbib\@empty
\bibitem [{\citenamefont {Lenarczyk}\ and\ \citenamefont
  {Luisier}(2016)}]{ref_myFeFET}%
  \BibitemOpen
  \bibfield  {author} {\bibinfo {author} {\bibfnamefont {P.}~\bibnamefont
  {Lenarczyk}}\ and\ \bibinfo {author} {\bibfnamefont {M.}~\bibnamefont
  {Luisier}},\ }in\ \href {\doibase 10.1109/SISPAD.2016.7605209} {\emph
  {\bibinfo {booktitle} {2016 International Conference on Simulation of
  Semiconductor Processes and Devices (SISPAD)}}}\ (\bibinfo {year} {2016})\
  pp.\ \bibinfo {pages} {311--314}\BibitemShut {NoStop}%
\bibitem [{\citenamefont {Resta}\ and\ \citenamefont
  {Vanderbilt}(2007)}]{ref_fieldTheory}%
  \BibitemOpen
  \bibfield  {author} {\bibinfo {author} {\bibfnamefont {R.}~\bibnamefont
  {Resta}}\ and\ \bibinfo {author} {\bibfnamefont {D.}~\bibnamefont
  {Vanderbilt}},\ }\enquote {\bibinfo {title} {Theory of polarization: A modern
  approach},}\ in\ \href {\doibase 10.1007/978-3-540-34591-6_2} {\emph
  {\bibinfo {booktitle} {Physics of Ferroelectrics: A Modern Perspective}}}\
  (\bibinfo  {publisher} {Springer Berlin Heidelberg},\ \bibinfo {address}
  {Berlin, Heidelberg},\ \bibinfo {year} {2007})\ pp.\ \bibinfo {pages}
  {31--68}\BibitemShut {NoStop}%
\bibitem [{\citenamefont {Giannozzi}\ \emph {et~al.}(1991)\citenamefont
  {Giannozzi}, \citenamefont {de~Gironcoli}, \citenamefont {Pavone},\ and\
  \citenamefont {Baroni}}]{ref_LRT}%
  \BibitemOpen
  \bibfield  {author} {\bibinfo {author} {\bibfnamefont {P.}~\bibnamefont
  {Giannozzi}}, \bibinfo {author} {\bibfnamefont {S.}~\bibnamefont
  {de~Gironcoli}}, \bibinfo {author} {\bibfnamefont {P.}~\bibnamefont
  {Pavone}}, \ and\ \bibinfo {author} {\bibfnamefont {S.}~\bibnamefont
  {Baroni}},\ }\href {\doibase 10.1103/PhysRevB.43.7231} {\bibfield  {journal}
  {\bibinfo  {journal} {Phys. Rev. B}\ }\textbf {\bibinfo {volume} {43}},\
  \bibinfo {pages} {7231} (\bibinfo {year} {1991})}\BibitemShut {NoStop}%
\bibitem [{\citenamefont {Levine}(1990)}]{ref_LRTdiv}%
  \BibitemOpen
  \bibfield  {author} {\bibinfo {author} {\bibfnamefont {Z.~H.}\ \bibnamefont
  {Levine}},\ }\href {\doibase 10.1103/PhysRevB.42.3567} {\bibfield  {journal}
  {\bibinfo  {journal} {Phys. Rev. B}\ }\textbf {\bibinfo {volume} {42}},\
  \bibinfo {pages} {3567} (\bibinfo {year} {1990})}\BibitemShut {NoStop}%
\bibitem [{\citenamefont {Nunes}\ and\ \citenamefont
  {Vanderbilt}(1994{\natexlab{a}})}]{ref_NV}%
  \BibitemOpen
  \bibfield  {author} {\bibinfo {author} {\bibfnamefont {R.~W.}\ \bibnamefont
  {Nunes}}\ and\ \bibinfo {author} {\bibfnamefont {D.}~\bibnamefont
  {Vanderbilt}},\ }\href {\doibase 10.1103/PhysRevLett.73.712} {\bibfield
  {journal} {\bibinfo  {journal} {Phys. Rev. Lett.}\ }\textbf {\bibinfo
  {volume} {73}},\ \bibinfo {pages} {712} (\bibinfo {year}
  {1994}{\natexlab{a}})}\BibitemShut {NoStop}%
\bibitem [{\citenamefont {King-Smith}\ and\ \citenamefont
  {Vanderbilt}(1993)}]{ref_MTP}%
  \BibitemOpen
  \bibfield  {author} {\bibinfo {author} {\bibfnamefont {R.~D.}\ \bibnamefont
  {King-Smith}}\ and\ \bibinfo {author} {\bibfnamefont {D.}~\bibnamefont
  {Vanderbilt}},\ }\href {\doibase 10.1103/PhysRevB.47.1651} {\bibfield
  {journal} {\bibinfo  {journal} {Phys. Rev. B}\ }\textbf {\bibinfo {volume}
  {47}},\ \bibinfo {pages} {1651} (\bibinfo {year} {1993})}\BibitemShut
  {NoStop}%
\bibitem [{\citenamefont {Fern\'andez}\ \emph {et~al.}(1998)\citenamefont
  {Fern\'andez}, \citenamefont {Dal~Corso},\ and\ \citenamefont
  {Baldereschi}}]{ref_polWf}%
  \BibitemOpen
  \bibfield  {author} {\bibinfo {author} {\bibfnamefont {P.}~\bibnamefont
  {Fern\'andez}}, \bibinfo {author} {\bibfnamefont {A.}~\bibnamefont
  {Dal~Corso}}, \ and\ \bibinfo {author} {\bibfnamefont {A.}~\bibnamefont
  {Baldereschi}},\ }\href {\doibase 10.1103/PhysRevB.58.R7480} {\bibfield
  {journal} {\bibinfo  {journal} {Phys. Rev. B}\ }\textbf {\bibinfo {volume}
  {58}},\ \bibinfo {pages} {R7480} (\bibinfo {year} {1998})}\BibitemShut
  {NoStop}%
\bibitem [{\citenamefont {Anderson}(1968)}]{ref_AndersonFunc}%
  \BibitemOpen
  \bibfield  {author} {\bibinfo {author} {\bibfnamefont {P.~W.}\ \bibnamefont
  {Anderson}},\ }\href {\doibase 10.1103/PhysRevLett.21.13} {\bibfield
  {journal} {\bibinfo  {journal} {Phys. Rev. Lett.}\ }\textbf {\bibinfo
  {volume} {21}},\ \bibinfo {pages} {13} (\bibinfo {year} {1968})}\BibitemShut
  {NoStop}%
\bibitem [{\citenamefont {Mortensen}\ and\ \citenamefont
  {Parrinello}(2001)}]{ref_orbSilicon}%
  \BibitemOpen
  \bibfield  {author} {\bibinfo {author} {\bibfnamefont {J.~J.}\ \bibnamefont
  {Mortensen}}\ and\ \bibinfo {author} {\bibfnamefont {M.}~\bibnamefont
  {Parrinello}},\ }\href {http://stacks.iop.org/0953-8984/13/i=25/a=301}
  {\bibfield  {journal} {\bibinfo  {journal} {J. Phys. Condens. Matter}\
  }\textbf {\bibinfo {volume} {13}},\ \bibinfo {pages} {5731} (\bibinfo {year}
  {2001})}\BibitemShut {NoStop}%
\bibitem [{\citenamefont {Hern\'andez}\ \emph {et~al.}(1996)\citenamefont
  {Hern\'andez}, \citenamefont {Gillan},\ and\ \citenamefont
  {Goringe}}]{ref_DM_ON}%
  \BibitemOpen
  \bibfield  {author} {\bibinfo {author} {\bibfnamefont {E.}~\bibnamefont
  {Hern\'andez}}, \bibinfo {author} {\bibfnamefont {M.~J.}\ \bibnamefont
  {Gillan}}, \ and\ \bibinfo {author} {\bibfnamefont {C.~M.}\ \bibnamefont
  {Goringe}},\ }\href {\doibase 10.1103/PhysRevB.53.7147} {\bibfield  {journal}
  {\bibinfo  {journal} {Phys. Rev. B}\ }\textbf {\bibinfo {volume} {53}},\
  \bibinfo {pages} {7147} (\bibinfo {year} {1996})}\BibitemShut {NoStop}%
\bibitem [{\citenamefont {Skylaris}\ \emph {et~al.}(2002)\citenamefont
  {Skylaris}, \citenamefont {Mostofi}, \citenamefont {Haynes}, \citenamefont
  {Di\'eguez},\ and\ \citenamefont {Payne}}]{ref_NGWF}%
  \BibitemOpen
  \bibfield  {author} {\bibinfo {author} {\bibfnamefont {C.-K.}\ \bibnamefont
  {Skylaris}}, \bibinfo {author} {\bibfnamefont {A.~A.}\ \bibnamefont
  {Mostofi}}, \bibinfo {author} {\bibfnamefont {P.~D.}\ \bibnamefont {Haynes}},
  \bibinfo {author} {\bibfnamefont {O.}~\bibnamefont {Di\'eguez}}, \ and\
  \bibinfo {author} {\bibfnamefont {M.~C.}\ \bibnamefont {Payne}},\ }\href
  {\doibase 10.1103/PhysRevB.66.035119} {\bibfield  {journal} {\bibinfo
  {journal} {Phys. Rev. B}\ }\textbf {\bibinfo {volume} {66}},\ \bibinfo
  {pages} {035119} (\bibinfo {year} {2002})}\BibitemShut {NoStop}%
\bibitem [{\citenamefont {Troullier}\ and\ \citenamefont
  {Martins}(1991)}]{ref_TMpseudo}%
  \BibitemOpen
  \bibfield  {author} {\bibinfo {author} {\bibfnamefont {N.}~\bibnamefont
  {Troullier}}\ and\ \bibinfo {author} {\bibfnamefont {J.~L.}\ \bibnamefont
  {Martins}},\ }\href {\doibase 10.1103/PhysRevB.43.1993} {\bibfield  {journal}
  {\bibinfo  {journal} {Phys. Rev. B}\ }\textbf {\bibinfo {volume} {43}},\
  \bibinfo {pages} {1993} (\bibinfo {year} {1991})}\BibitemShut {NoStop}%
\bibitem [{\citenamefont {Kohn}\ and\ \citenamefont {Sham}(1965)}]{ref_KS}%
  \BibitemOpen
  \bibfield  {author} {\bibinfo {author} {\bibfnamefont {W.}~\bibnamefont
  {Kohn}}\ and\ \bibinfo {author} {\bibfnamefont {L.~J.}\ \bibnamefont
  {Sham}},\ }\href {\doibase 10.1103/PhysRev.140.A1133} {\bibfield  {journal}
  {\bibinfo  {journal} {Phys. Rev.}\ }\textbf {\bibinfo {volume} {140}},\
  \bibinfo {pages} {A1133} (\bibinfo {year} {1965})}\BibitemShut {NoStop}%
\bibitem [{\citenamefont {Hohenberg}\ and\ \citenamefont
  {Kohn}(1964)}]{ref_HK}%
  \BibitemOpen
  \bibfield  {author} {\bibinfo {author} {\bibfnamefont {P.}~\bibnamefont
  {Hohenberg}}\ and\ \bibinfo {author} {\bibfnamefont {W.}~\bibnamefont
  {Kohn}},\ }\href {\doibase 10.1103/PhysRev.136.B864} {\bibfield  {journal}
  {\bibinfo  {journal} {Phys. Rev.}\ }\textbf {\bibinfo {volume} {136}},\
  \bibinfo {pages} {B864} (\bibinfo {year} {1964})}\BibitemShut {NoStop}%
\bibitem [{\citenamefont {Payne}\ \emph {et~al.}(1992)\citenamefont {Payne},
  \citenamefont {Teter}, \citenamefont {Allan}, \citenamefont {Arias},\ and\
  \citenamefont {Joannopoulos}}]{ref_iterMin}%
  \BibitemOpen
  \bibfield  {author} {\bibinfo {author} {\bibfnamefont {M.~C.}\ \bibnamefont
  {Payne}}, \bibinfo {author} {\bibfnamefont {M.~P.}\ \bibnamefont {Teter}},
  \bibinfo {author} {\bibfnamefont {D.~C.}\ \bibnamefont {Allan}}, \bibinfo
  {author} {\bibfnamefont {T.~A.}\ \bibnamefont {Arias}}, \ and\ \bibinfo
  {author} {\bibfnamefont {J.~D.}\ \bibnamefont {Joannopoulos}},\ }\href
  {\doibase 10.1103/RevModPhys.64.1045} {\bibfield  {journal} {\bibinfo
  {journal} {Rev. Mod. Phys.}\ }\textbf {\bibinfo {volume} {64}},\ \bibinfo
  {pages} {1045} (\bibinfo {year} {1992})}\BibitemShut {NoStop}%
\bibitem [{\citenamefont {Marzari}\ \emph {et~al.}(2012)\citenamefont
  {Marzari}, \citenamefont {Mostofi}, \citenamefont {Yates}, \citenamefont
  {Souza},\ and\ \citenamefont {Vanderbilt}}]{ref_WFrev}%
  \BibitemOpen
  \bibfield  {author} {\bibinfo {author} {\bibfnamefont {N.}~\bibnamefont
  {Marzari}}, \bibinfo {author} {\bibfnamefont {A.~A.}\ \bibnamefont
  {Mostofi}}, \bibinfo {author} {\bibfnamefont {J.~R.}\ \bibnamefont {Yates}},
  \bibinfo {author} {\bibfnamefont {I.}~\bibnamefont {Souza}}, \ and\ \bibinfo
  {author} {\bibfnamefont {D.}~\bibnamefont {Vanderbilt}},\ }\href {\doibase
  10.1103/RevModPhys.84.1419} {\bibfield  {journal} {\bibinfo  {journal} {Rev.
  Mod. Phys.}\ }\textbf {\bibinfo {volume} {84}},\ \bibinfo {pages} {1419}
  (\bibinfo {year} {2012})}\BibitemShut {NoStop}%
\bibitem [{\citenamefont {Hierse}\ and\ \citenamefont
  {Stechel}(1994)}]{ref_DMparam}%
  \BibitemOpen
  \bibfield  {author} {\bibinfo {author} {\bibfnamefont {W.}~\bibnamefont
  {Hierse}}\ and\ \bibinfo {author} {\bibfnamefont {E.~B.}\ \bibnamefont
  {Stechel}},\ }\href {\doibase 10.1103/PhysRevB.50.17811} {\bibfield
  {journal} {\bibinfo  {journal} {Phys. Rev. B}\ }\textbf {\bibinfo {volume}
  {50}},\ \bibinfo {pages} {17811} (\bibinfo {year} {1994})}\BibitemShut
  {NoStop}%
\bibitem [{\citenamefont {Stechel}\ \emph {et~al.}(1994)\citenamefont
  {Stechel}, \citenamefont {Williams},\ and\ \citenamefont
  {Feibelman}}]{ref_ONmetals}%
  \BibitemOpen
  \bibfield  {author} {\bibinfo {author} {\bibfnamefont {E.~B.}\ \bibnamefont
  {Stechel}}, \bibinfo {author} {\bibfnamefont {A.~R.}\ \bibnamefont
  {Williams}}, \ and\ \bibinfo {author} {\bibfnamefont {P.~J.}\ \bibnamefont
  {Feibelman}},\ }\href {\doibase 10.1103/PhysRevB.49.10088} {\bibfield
  {journal} {\bibinfo  {journal} {Phys. Rev. B}\ }\textbf {\bibinfo {volume}
  {49}},\ \bibinfo {pages} {10088} (\bibinfo {year} {1994})}\BibitemShut
  {NoStop}%
\bibitem [{\citenamefont {McWeeny}(1960)}]{ref_McWeeny}%
  \BibitemOpen
  \bibfield  {author} {\bibinfo {author} {\bibfnamefont {R.}~\bibnamefont
  {McWeeny}},\ }\href {\doibase 10.1103/RevModPhys.32.335} {\bibfield
  {journal} {\bibinfo  {journal} {Rev. Mod. Phys.}\ }\textbf {\bibinfo {volume}
  {32}},\ \bibinfo {pages} {335} (\bibinfo {year} {1960})}\BibitemShut
  {NoStop}%
\bibitem [{\citenamefont {Galli}(1996)}]{ref_ONrev}%
  \BibitemOpen
  \bibfield  {author} {\bibinfo {author} {\bibfnamefont {G.}~\bibnamefont
  {Galli}},\ }\href {\doibase https://doi.org/10.1016/S1359-0286(96)80114-8}
  {\bibfield  {journal} {\bibinfo  {journal} {Curr. Opin. Solid State Mater.
  Sci.}\ }\textbf {\bibinfo {volume} {1}},\ \bibinfo {pages} {864 } (\bibinfo
  {year} {1996})}\BibitemShut {NoStop}%
\bibitem [{\citenamefont {Chelikowsky}()}]{ref_PARSEC_web}%
  \BibitemOpen
  \bibfield  {author} {\bibinfo {author} {\bibfnamefont {J.~R.}\ \bibnamefont
  {Chelikowsky}},\ }\href@noop {} {\enquote {\bibinfo {title} {{PARSEC} quantum
  mechanics applied to materials},}\ }\bibinfo {howpublished}
  {\url{http://parsec.ices.utexas.edu/}},\ \bibinfo {note} {accessed:
  2018-05-15}\BibitemShut {NoStop}%
\bibitem [{\citenamefont {Chelikowsky}\ \emph {et~al.}(1994)\citenamefont
  {Chelikowsky}, \citenamefont {Troullier},\ and\ \citenamefont
  {Saad}}]{ref_PARSEC_FD}%
  \BibitemOpen
  \bibfield  {author} {\bibinfo {author} {\bibfnamefont {J.~R.}\ \bibnamefont
  {Chelikowsky}}, \bibinfo {author} {\bibfnamefont {N.}~\bibnamefont
  {Troullier}}, \ and\ \bibinfo {author} {\bibfnamefont {Y.}~\bibnamefont
  {Saad}},\ }\href {\doibase 10.1103/PhysRevLett.72.1240} {\bibfield  {journal}
  {\bibinfo  {journal} {Phys. Rev. Lett.}\ }\textbf {\bibinfo {volume} {72}},\
  \bibinfo {pages} {1240} (\bibinfo {year} {1994})}\BibitemShut {NoStop}%
\bibitem [{\citenamefont {Kleinman}\ and\ \citenamefont
  {Bylander}(1982)}]{ref_KBpseudo}%
  \BibitemOpen
  \bibfield  {author} {\bibinfo {author} {\bibfnamefont {L.}~\bibnamefont
  {Kleinman}}\ and\ \bibinfo {author} {\bibfnamefont {D.~M.}\ \bibnamefont
  {Bylander}},\ }\href {\doibase 10.1103/PhysRevLett.48.1425} {\bibfield
  {journal} {\bibinfo  {journal} {Phys. Rev. Lett.}\ }\textbf {\bibinfo
  {volume} {48}},\ \bibinfo {pages} {1425} (\bibinfo {year}
  {1982})}\BibitemShut {NoStop}%
\bibitem [{\citenamefont {Ceperley}\ and\ \citenamefont
  {Alder}(1980)}]{ref_LDA_CA}%
  \BibitemOpen
  \bibfield  {author} {\bibinfo {author} {\bibfnamefont {D.~M.}\ \bibnamefont
  {Ceperley}}\ and\ \bibinfo {author} {\bibfnamefont {B.~J.}\ \bibnamefont
  {Alder}},\ }\href {\doibase 10.1103/PhysRevLett.45.566} {\bibfield  {journal}
  {\bibinfo  {journal} {Phys. Rev. Lett.}\ }\textbf {\bibinfo {volume} {45}},\
  \bibinfo {pages} {566} (\bibinfo {year} {1980})}\BibitemShut {NoStop}%
\bibitem [{\citenamefont {Perdew}\ and\ \citenamefont
  {Zunger}(1981)}]{ref_LDA_PZ}%
  \BibitemOpen
  \bibfield  {author} {\bibinfo {author} {\bibfnamefont {J.~P.}\ \bibnamefont
  {Perdew}}\ and\ \bibinfo {author} {\bibfnamefont {A.}~\bibnamefont
  {Zunger}},\ }\href {\doibase 10.1103/PhysRevB.23.5048} {\bibfield  {journal}
  {\bibinfo  {journal} {Phys. Rev. B}\ }\textbf {\bibinfo {volume} {23}},\
  \bibinfo {pages} {5048} (\bibinfo {year} {1981})}\BibitemShut {NoStop}%
\bibitem [{\citenamefont {Polak}(1971)}]{ref_CG_PR}%
  \BibitemOpen
  \bibfield  {author} {\bibinfo {author} {\bibfnamefont {E.}~\bibnamefont
  {Polak}},\ }\href@noop {} {\emph {\bibinfo {title} {Computational methods in
  optimization : a unified approach}}},\ \bibinfo {series} {Mathematics in
  science and engineering}, Vol.~\bibinfo {volume} {77}\ (\bibinfo  {publisher}
  {Academic Press},\ \bibinfo {address} {New York},\ \bibinfo {year}
  {1971})\BibitemShut {NoStop}%
\bibitem [{\citenamefont {Haynes}\ \emph {et~al.}(2008)\citenamefont {Haynes},
  \citenamefont {Skylaris}, \citenamefont {Mostofi},\ and\ \citenamefont
  {Payne}}]{ref_ONETEP_Kmin}%
  \BibitemOpen
  \bibfield  {author} {\bibinfo {author} {\bibfnamefont {P.~D.}\ \bibnamefont
  {Haynes}}, \bibinfo {author} {\bibfnamefont {C.-K.}\ \bibnamefont
  {Skylaris}}, \bibinfo {author} {\bibfnamefont {A.~A.}\ \bibnamefont
  {Mostofi}}, \ and\ \bibinfo {author} {\bibfnamefont {M.~C.}\ \bibnamefont
  {Payne}},\ }\href {http://stacks.iop.org/0953-8984/20/i=29/a=294207}
  {\bibfield  {journal} {\bibinfo  {journal} {J. Phys. Condens. Matter}\
  }\textbf {\bibinfo {volume} {20}},\ \bibinfo {pages} {294207} (\bibinfo
  {year} {2008})}\BibitemShut {NoStop}%
\bibitem [{\citenamefont {Press}\ \emph {et~al.}(2007)\citenamefont {Press},
  \citenamefont {Teukolsky}, \citenamefont {Vetterling},\ and\ \citenamefont
  {Flannery}}]{ref_NumRecipes}%
  \BibitemOpen
  \bibfield  {author} {\bibinfo {author} {\bibfnamefont {W.~H.}\ \bibnamefont
  {Press}}, \bibinfo {author} {\bibfnamefont {S.~A.}\ \bibnamefont
  {Teukolsky}}, \bibinfo {author} {\bibfnamefont {W.~T.}\ \bibnamefont
  {Vetterling}}, \ and\ \bibinfo {author} {\bibfnamefont {B.~P.}\ \bibnamefont
  {Flannery}},\ }\href@noop {} {\emph {\bibinfo {title} {Numerical Recipes 3rd
  Edition: The Art of Scientific Computing}}},\ \bibinfo {edition} {3rd}\ ed.\
  (\bibinfo  {publisher} {Cambridge University Press},\ \bibinfo {address} {New
  York, NY, USA},\ \bibinfo {year} {2007})\BibitemShut {NoStop}%
\bibitem [{\citenamefont {Martin}(2004)}]{ref_elecSt}%
  \BibitemOpen
  \bibfield  {author} {\bibinfo {author} {\bibfnamefont {R.~M.}\ \bibnamefont
  {Martin}},\ }\href {\doibase 10.1017/CBO9780511805769} {\emph {\bibinfo
  {title} {Electronic Structure: Basic Theory and Practical Methods}}}\
  (\bibinfo  {publisher} {Cambridge University Press},\ \bibinfo {year}
  {2004})\BibitemShut {NoStop}%
\bibitem [{\citenamefont {Cloizeaux}(1964)}]{ref_projAnalytic}%
  \BibitemOpen
  \bibfield  {author} {\bibinfo {author} {\bibfnamefont {J.~D.}\ \bibnamefont
  {Cloizeaux}},\ }\href {\doibase 10.1103/PhysRev.135.A685} {\bibfield
  {journal} {\bibinfo  {journal} {Phys. Rev.}\ }\textbf {\bibinfo {volume}
  {135}},\ \bibinfo {pages} {A685} (\bibinfo {year} {1964})}\BibitemShut
  {NoStop}%
\bibitem [{\citenamefont {Kohn}(1959)}]{ref_expDecay_Kohn}%
  \BibitemOpen
  \bibfield  {author} {\bibinfo {author} {\bibfnamefont {W.}~\bibnamefont
  {Kohn}},\ }\href {\doibase 10.1103/PhysRev.115.809} {\bibfield  {journal}
  {\bibinfo  {journal} {Phys. Rev.}\ }\textbf {\bibinfo {volume} {115}},\
  \bibinfo {pages} {809} (\bibinfo {year} {1959})}\BibitemShut {NoStop}%
\bibitem [{\citenamefont {He}\ and\ \citenamefont
  {Vanderbilt}(2001)}]{ref_expDecay_HV}%
  \BibitemOpen
  \bibfield  {author} {\bibinfo {author} {\bibfnamefont {L.}~\bibnamefont
  {He}}\ and\ \bibinfo {author} {\bibfnamefont {D.}~\bibnamefont
  {Vanderbilt}},\ }\href {\doibase 10.1103/PhysRevLett.86.5341} {\bibfield
  {journal} {\bibinfo  {journal} {Phys. Rev. Lett.}\ }\textbf {\bibinfo
  {volume} {86}},\ \bibinfo {pages} {5341} (\bibinfo {year}
  {2001})}\BibitemShut {NoStop}%
\bibitem [{\citenamefont {Nenciu}(1983)}]{ref_expLocWF}%
  \BibitemOpen
  \bibfield  {author} {\bibinfo {author} {\bibfnamefont {G.}~\bibnamefont
  {Nenciu}},\ }\href {\doibase 10.1007/BF01206052} {\bibfield  {journal}
  {\bibinfo  {journal} {Communications in Mathematical Physics}\ }\textbf
  {\bibinfo {volume} {91}},\ \bibinfo {pages} {81} (\bibinfo {year}
  {1983})}\BibitemShut {NoStop}%
\bibitem [{\citenamefont {Kohn}(1993)}]{ref_Kohn_DFWFT}%
  \BibitemOpen
  \bibfield  {author} {\bibinfo {author} {\bibfnamefont {W.}~\bibnamefont
  {Kohn}},\ }\href {\doibase https://doi.org/10.1016/0009-2614(93)89056-N}
  {\bibfield  {journal} {\bibinfo  {journal} {Chemical Physics Letters}\
  }\textbf {\bibinfo {volume} {208}},\ \bibinfo {pages} {167 } (\bibinfo {year}
  {1993})}\BibitemShut {NoStop}%
\bibitem [{\citenamefont {Ismail-Beigi}\ and\ \citenamefont
  {Arias}(1999)}]{ref_locDM_weakBind}%
  \BibitemOpen
  \bibfield  {author} {\bibinfo {author} {\bibfnamefont {S.}~\bibnamefont
  {Ismail-Beigi}}\ and\ \bibinfo {author} {\bibfnamefont {T.~A.}\ \bibnamefont
  {Arias}},\ }\href {\doibase 10.1103/PhysRevLett.82.2127} {\bibfield
  {journal} {\bibinfo  {journal} {Phys. Rev. Lett.}\ }\textbf {\bibinfo
  {volume} {82}},\ \bibinfo {pages} {2127} (\bibinfo {year}
  {1999})}\BibitemShut {NoStop}%
\bibitem [{\citenamefont {Niklasson}(2002)}]{ref_DMalg1}%
  \BibitemOpen
  \bibfield  {author} {\bibinfo {author} {\bibfnamefont {A.~M.~N.}\
  \bibnamefont {Niklasson}},\ }\href {\doibase 10.1103/PhysRevB.66.155115}
  {\bibfield  {journal} {\bibinfo  {journal} {Phys. Rev. B}\ }\textbf {\bibinfo
  {volume} {66}},\ \bibinfo {pages} {155115} (\bibinfo {year}
  {2002})}\BibitemShut {NoStop}%
\bibitem [{\citenamefont {Niklasson}\ and\ \citenamefont
  {Challacombe}(2004)}]{ref_DMalg2}%
  \BibitemOpen
  \bibfield  {author} {\bibinfo {author} {\bibfnamefont {A.~M.~N.}\
  \bibnamefont {Niklasson}}\ and\ \bibinfo {author} {\bibfnamefont
  {M.}~\bibnamefont {Challacombe}},\ }\href {\doibase
  10.1103/PhysRevLett.92.193001} {\bibfield  {journal} {\bibinfo  {journal}
  {Phys. Rev. Lett.}\ }\textbf {\bibinfo {volume} {92}},\ \bibinfo {pages}
  {193001} (\bibinfo {year} {2004})}\BibitemShut {NoStop}%
\bibitem [{\citenamefont {Niklasson}\ \emph {et~al.}(2003)\citenamefont
  {Niklasson}, \citenamefont {Tymczak},\ and\ \citenamefont
  {Challacombe}}]{ref_DMalg3}%
  \BibitemOpen
  \bibfield  {author} {\bibinfo {author} {\bibfnamefont {A.~M.~N.}\
  \bibnamefont {Niklasson}}, \bibinfo {author} {\bibfnamefont {C.~J.}\
  \bibnamefont {Tymczak}}, \ and\ \bibinfo {author} {\bibfnamefont
  {M.}~\bibnamefont {Challacombe}},\ }\href {\doibase 10.1063/1.1559913}
  {\bibfield  {journal} {\bibinfo  {journal} {The Journal of Chemical Physics}\
  }\textbf {\bibinfo {volume} {118}},\ \bibinfo {pages} {8611} (\bibinfo {year}
  {2003})}\BibitemShut {NoStop}%
\bibitem [{\citenamefont {Niklasson}\ \emph {et~al.}(2005)\citenamefont
  {Niklasson}, \citenamefont {Weber},\ and\ \citenamefont
  {Challacombe}}]{ref_DMalg4}%
  \BibitemOpen
  \bibfield  {author} {\bibinfo {author} {\bibfnamefont {A.~M.~N.}\
  \bibnamefont {Niklasson}}, \bibinfo {author} {\bibfnamefont {V.}~\bibnamefont
  {Weber}}, \ and\ \bibinfo {author} {\bibfnamefont {M.}~\bibnamefont
  {Challacombe}},\ }\href {\doibase 10.1063/1.1944725} {\bibfield  {journal}
  {\bibinfo  {journal} {The Journal of Chemical Physics}\ }\textbf {\bibinfo
  {volume} {123}},\ \bibinfo {pages} {044107} (\bibinfo {year}
  {2005})}\BibitemShut {NoStop}%
\bibitem [{\citenamefont {Li}\ \emph {et~al.}(1993)\citenamefont {Li},
  \citenamefont {Nunes},\ and\ \citenamefont {Vanderbilt}}]{ref_DM_LNV}%
  \BibitemOpen
  \bibfield  {author} {\bibinfo {author} {\bibfnamefont {X.-P.}\ \bibnamefont
  {Li}}, \bibinfo {author} {\bibfnamefont {R.~W.}\ \bibnamefont {Nunes}}, \
  and\ \bibinfo {author} {\bibfnamefont {D.}~\bibnamefont {Vanderbilt}},\
  }\href {\doibase 10.1103/PhysRevB.47.10891} {\bibfield  {journal} {\bibinfo
  {journal} {Phys. Rev. B}\ }\textbf {\bibinfo {volume} {47}},\ \bibinfo
  {pages} {10891} (\bibinfo {year} {1993})}\BibitemShut {NoStop}%
\bibitem [{\citenamefont {Nunes}\ and\ \citenamefont
  {Vanderbilt}(1994{\natexlab{b}})}]{ref_DM_NV}%
  \BibitemOpen
  \bibfield  {author} {\bibinfo {author} {\bibfnamefont {R.~W.}\ \bibnamefont
  {Nunes}}\ and\ \bibinfo {author} {\bibfnamefont {D.}~\bibnamefont
  {Vanderbilt}},\ }\href {\doibase 10.1103/PhysRevB.50.17611} {\bibfield
  {journal} {\bibinfo  {journal} {Phys. Rev. B}\ }\textbf {\bibinfo {volume}
  {50}},\ \bibinfo {pages} {17611} (\bibinfo {year}
  {1994}{\natexlab{b}})}\BibitemShut {NoStop}%
\bibitem [{\citenamefont {Marzari}\ and\ \citenamefont
  {Vanderbilt}(1998)}]{ref_btoWf}%
  \BibitemOpen
  \bibfield  {author} {\bibinfo {author} {\bibfnamefont {N.}~\bibnamefont
  {Marzari}}\ and\ \bibinfo {author} {\bibfnamefont {D.}~\bibnamefont
  {Vanderbilt}},\ }\href@noop {} {\bibfield  {journal} {\bibinfo  {journal}
  {First-Principles Calculations for Ferroelectrics}\ }\bibinfo {series} {Aip
  Conference Proceedings},\ \bibinfo {pages} {146} (\bibinfo {year}
  {1998})}\BibitemShut {NoStop}%
\bibitem [{\citenamefont {Monkhorst}\ and\ \citenamefont
  {Pack}(1976)}]{ref_MPgrid}%
  \BibitemOpen
  \bibfield  {author} {\bibinfo {author} {\bibfnamefont {H.~J.}\ \bibnamefont
  {Monkhorst}}\ and\ \bibinfo {author} {\bibfnamefont {J.~D.}\ \bibnamefont
  {Pack}},\ }\href {\doibase 10.1103/PhysRevB.13.5188} {\bibfield  {journal}
  {\bibinfo  {journal} {Phys. Rev. B}\ }\textbf {\bibinfo {volume} {13}},\
  \bibinfo {pages} {5188} (\bibinfo {year} {1976})}\BibitemShut {NoStop}%
\bibitem [{\citenamefont {Dal~Corso}\ \emph {et~al.}(1994)\citenamefont
  {Dal~Corso}, \citenamefont {Baroni},\ and\ \citenamefont
  {Resta}}]{ref_epsSiLRT_1}%
  \BibitemOpen
  \bibfield  {author} {\bibinfo {author} {\bibfnamefont {A.}~\bibnamefont
  {Dal~Corso}}, \bibinfo {author} {\bibfnamefont {S.}~\bibnamefont {Baroni}}, \
  and\ \bibinfo {author} {\bibfnamefont {R.}~\bibnamefont {Resta}},\ }\href
  {\doibase 10.1103/PhysRevB.49.5323} {\bibfield  {journal} {\bibinfo
  {journal} {Phys. Rev. B}\ }\textbf {\bibinfo {volume} {49}},\ \bibinfo
  {pages} {5323} (\bibinfo {year} {1994})}\BibitemShut {NoStop}%
\bibitem [{\citenamefont {Levine}\ and\ \citenamefont
  {Allan}(1991)}]{ref_epsSiLRT_2}%
  \BibitemOpen
  \bibfield  {author} {\bibinfo {author} {\bibfnamefont {Z.~H.}\ \bibnamefont
  {Levine}}\ and\ \bibinfo {author} {\bibfnamefont {D.~C.}\ \bibnamefont
  {Allan}},\ }\href {\doibase 10.1103/PhysRevB.44.12781} {\bibfield  {journal}
  {\bibinfo  {journal} {Phys. Rev. B}\ }\textbf {\bibinfo {volume} {44}},\
  \bibinfo {pages} {12781} (\bibinfo {year} {1991})}\BibitemShut {NoStop}%
\bibitem [{\citenamefont {Momma}\ and\ \citenamefont
  {Izumi}(2011)}]{ref_Vesta}%
  \BibitemOpen
  \bibfield  {author} {\bibinfo {author} {\bibfnamefont {K.}~\bibnamefont
  {Momma}}\ and\ \bibinfo {author} {\bibfnamefont {F.}~\bibnamefont {Izumi}},\
  }\href {\doibase 10.1107/S0021889811038970} {\bibfield  {journal} {\bibinfo
  {journal} {J. Appl. Crystallogr.}\ }\textbf {\bibinfo {volume} {44}},\
  \bibinfo {pages} {1272} (\bibinfo {year} {2011})}\BibitemShut {NoStop}%
\bibitem [{\citenamefont {Ghosez}\ \emph {et~al.}(1998)\citenamefont {Ghosez},
  \citenamefont {Gonze},\ and\ \citenamefont {Michenaud}}]{Ref_epsBTO_LRT}%
  \BibitemOpen
  \bibfield  {author} {\bibinfo {author} {\bibfnamefont {P.~H.}\ \bibnamefont
  {Ghosez}}, \bibinfo {author} {\bibfnamefont {X.}~\bibnamefont {Gonze}}, \
  and\ \bibinfo {author} {\bibfnamefont {J.~P.}\ \bibnamefont {Michenaud}},\
  }\href {\doibase 10.1080/00150199808009159} {\bibfield  {journal} {\bibinfo
  {journal} {Ferroelectrics}\ }\textbf {\bibinfo {volume} {206}},\ \bibinfo
  {pages} {205} (\bibinfo {year} {1998})}\BibitemShut {NoStop}%
\bibitem [{\citenamefont {Axe}(1967)}]{Ref_epsBTO_exp}%
  \BibitemOpen
  \bibfield  {author} {\bibinfo {author} {\bibfnamefont {J.~D.}\ \bibnamefont
  {Axe}},\ }\href {\doibase 10.1103/PhysRev.157.429} {\bibfield  {journal}
  {\bibinfo  {journal} {Phys. Rev.}\ }\textbf {\bibinfo {volume} {157}},\
  \bibinfo {pages} {429} (\bibinfo {year} {1967})}\BibitemShut {NoStop}%
\bibitem [{\citenamefont {Posternak}\ \emph {et~al.}(1994)\citenamefont
  {Posternak}, \citenamefont {Resta},\ and\ \citenamefont
  {Baldereschi}}]{ref_ferroOrigin}%
  \BibitemOpen
  \bibfield  {author} {\bibinfo {author} {\bibfnamefont {M.}~\bibnamefont
  {Posternak}}, \bibinfo {author} {\bibfnamefont {R.}~\bibnamefont {Resta}}, \
  and\ \bibinfo {author} {\bibfnamefont {A.}~\bibnamefont {Baldereschi}},\
  }\href {\doibase 10.1103/PhysRevB.50.8911} {\bibfield  {journal} {\bibinfo
  {journal} {Phys. Rev. B}\ }\textbf {\bibinfo {volume} {50}},\ \bibinfo
  {pages} {8911} (\bibinfo {year} {1994})}\BibitemShut {NoStop}%
\bibitem [{\citenamefont {Kim}\ \emph {et~al.}(1995)\citenamefont {Kim},
  \citenamefont {Mauri},\ and\ \citenamefont {Galli}}]{ref_KMG}%
  \BibitemOpen
  \bibfield  {author} {\bibinfo {author} {\bibfnamefont {J.}~\bibnamefont
  {Kim}}, \bibinfo {author} {\bibfnamefont {F.}~\bibnamefont {Mauri}}, \ and\
  \bibinfo {author} {\bibfnamefont {G.}~\bibnamefont {Galli}},\ }\href
  {\doibase 10.1103/PhysRevB.52.1640} {\bibfield  {journal} {\bibinfo
  {journal} {Phys. Rev. B}\ }\textbf {\bibinfo {volume} {52}},\ \bibinfo
  {pages} {1640} (\bibinfo {year} {1995})}\BibitemShut {NoStop}%
\bibitem [{\citenamefont {Galli}\ and\ \citenamefont
  {Parrinello}(1992)}]{ref_largeScale_overlapping}%
  \BibitemOpen
  \bibfield  {author} {\bibinfo {author} {\bibfnamefont {G.}~\bibnamefont
  {Galli}}\ and\ \bibinfo {author} {\bibfnamefont {M.}~\bibnamefont
  {Parrinello}},\ }\href {\doibase 10.1103/PhysRevLett.69.3547} {\bibfield
  {journal} {\bibinfo  {journal} {Phys. Rev. Lett.}\ }\textbf {\bibinfo
  {volume} {69}},\ \bibinfo {pages} {3547} (\bibinfo {year}
  {1992})}\BibitemShut {NoStop}%
\bibitem [{\citenamefont {Arias}\ \emph {et~al.}(1992)\citenamefont {Arias},
  \citenamefont {Payne},\ and\ \citenamefont
  {Joannopoulos}}]{ref_MD_overlapping}%
  \BibitemOpen
  \bibfield  {author} {\bibinfo {author} {\bibfnamefont {T.~A.}\ \bibnamefont
  {Arias}}, \bibinfo {author} {\bibfnamefont {M.~C.}\ \bibnamefont {Payne}}, \
  and\ \bibinfo {author} {\bibfnamefont {J.~D.}\ \bibnamefont {Joannopoulos}},\
  }\href {\doibase 10.1103/PhysRevLett.69.1077} {\bibfield  {journal} {\bibinfo
   {journal} {Phys. Rev. Lett.}\ }\textbf {\bibinfo {volume} {69}},\ \bibinfo
  {pages} {1077} (\bibinfo {year} {1992})}\BibitemShut {NoStop}%
\bibitem [{\citenamefont {Mauri}\ \emph {et~al.}(1993)\citenamefont {Mauri},
  \citenamefont {Galli},\ and\ \citenamefont {Car}}]{ref_MGC}%
  \BibitemOpen
  \bibfield  {author} {\bibinfo {author} {\bibfnamefont {F.}~\bibnamefont
  {Mauri}}, \bibinfo {author} {\bibfnamefont {G.}~\bibnamefont {Galli}}, \ and\
  \bibinfo {author} {\bibfnamefont {R.}~\bibnamefont {Car}},\ }\href {\doibase
  10.1103/PhysRevB.47.9973} {\bibfield  {journal} {\bibinfo  {journal} {Phys.
  Rev. B}\ }\textbf {\bibinfo {volume} {47}},\ \bibinfo {pages} {9973}
  (\bibinfo {year} {1993})}\BibitemShut {NoStop}%
\bibitem [{\citenamefont {Horn}\ and\ \citenamefont
  {Johnson}(1985)}]{Ref_matrix_PD}%
  \BibitemOpen
  \bibfield  {author} {\bibinfo {author} {\bibfnamefont {R.~A.}\ \bibnamefont
  {Horn}}\ and\ \bibinfo {author} {\bibfnamefont {C.~R.}\ \bibnamefont
  {Johnson}},\ }\enquote {\bibinfo {title} {Positive definite matrices},}\ in\
  \href@noop {} {\emph {\bibinfo {booktitle} {Matrix Analysis}}}\ (\bibinfo
  {publisher} {Cambridge University Press},\ \bibinfo {year} {1985})\ pp.\
  \bibinfo {pages} {391--486}\BibitemShut {NoStop}%
\bibitem [{\citenamefont {Ordej\'on}\ \emph {et~al.}(1993)\citenamefont
  {Ordej\'on}, \citenamefont {Drabold}, \citenamefont {Grumbach},\ and\
  \citenamefont {Martin}}]{ref_Ordejon_unconstrained}%
  \BibitemOpen
  \bibfield  {author} {\bibinfo {author} {\bibfnamefont {P.}~\bibnamefont
  {Ordej\'on}}, \bibinfo {author} {\bibfnamefont {D.~A.}\ \bibnamefont
  {Drabold}}, \bibinfo {author} {\bibfnamefont {M.~P.}\ \bibnamefont
  {Grumbach}}, \ and\ \bibinfo {author} {\bibfnamefont {R.~M.}\ \bibnamefont
  {Martin}},\ }\href {\doibase 10.1103/PhysRevB.48.14646} {\bibfield  {journal}
  {\bibinfo  {journal} {Phys. Rev. B}\ }\textbf {\bibinfo {volume} {48}},\
  \bibinfo {pages} {14646} (\bibinfo {year} {1993})}\BibitemShut {NoStop}%
\bibitem [{\citenamefont {Resta}(1998)}]{ref_illPosedPosOp}%
  \BibitemOpen
  \bibfield  {author} {\bibinfo {author} {\bibfnamefont {R.}~\bibnamefont
  {Resta}},\ }\href {\doibase 10.1103/PhysRevLett.80.1800} {\bibfield
  {journal} {\bibinfo  {journal} {Phys. Rev. Lett.}\ }\textbf {\bibinfo
  {volume} {80}},\ \bibinfo {pages} {1800} (\bibinfo {year}
  {1998})}\BibitemShut {NoStop}%
\bibitem [{\citenamefont {Griffiths}(2017)}]{ref_QMintro}%
  \BibitemOpen
  \bibfield  {author} {\bibinfo {author} {\bibfnamefont {D.~J.}\ \bibnamefont
  {Griffiths}},\ }\href@noop {} {\emph {\bibinfo {title} {Introduction to
  quantum mechanics}}},\ \bibinfo {edition} {second edition}\ ed.\ (\bibinfo
  {publisher} {Cambridge University Press},\ \bibinfo {address} {Cambridge
  United Kingdom},\ \bibinfo {year} {2017})\BibitemShut {NoStop}%
\end{thebibliography}%

\end{document}